\documentclass[reprint,aps,prb,superscriptaddress,longbibliography,amsmath,amssymb,twocolumn]{revtex4-1}
\usepackage{graphicx}

\usepackage{xcolor}
\usepackage{mathtools}
\usepackage{bm}
\usepackage{bbm}
\usepackage[caption=false]{subfig}
\usepackage[colorlinks=true,allcolors=blue]{hyperref}
\usepackage{gensymb}
\usepackage{soul}
\usepackage{dsfont}
\usepackage{physics}
\usepackage{chngcntr}

\renewcommand\vec{\mathbf}
\newcommand\ddp{{\rm{d}}p\,}

\definecolor{TTH-color2}{named}{blue}
\definecolor{TH-color2}{named}{green}
\definecolor{RO-color2}{named}{red}
\definecolor{MS-color2}{RGB}{128,0,128}

\newcommand{\TCm}{{T_C^{\rm m}}}
\newcommand{\TCsc}{{T_C^{\rm sc}}}
\newcommand{\vF}{v_{\rm F}}

\begin{document}

\title{Competition of electron-phonon mediated superconductivity and Stoner 
magnetism on a flat band}

\author{Risto Ojaj\"arvi}
\affiliation{Department of Physics and Nanoscience Center, University of Jyvaskyla, P.O. Box 35 (YFL), FI-40014 University of Jyvaskyla, Finland}

\author{Timo Hyart}
\affiliation{Department of Physics and Nanoscience Center, University of Jyvaskyla, P.O. Box 35 (YFL), FI-40014 University of Jyvaskyla, Finland}
\affiliation{Institut f\"ur Theoretische Physik, Universit\"at Leipzig, D-04103 Leipzig, Germany}

\author{Mihail A. Silaev}
\author{Tero T.~Heikkil\"a}

\affiliation{Department of Physics and Nanoscience Center, University of Jyvaskyla, P.O. Box 35 (YFL), FI-40014 University of Jyvaskyla, Finland}

\date{\today}

\begin{abstract}
The effective attractive interaction between electrons, mediated by electron-phonon coupling, is a well-established mechanism of conventional superconductivity. In metals exhibiting a Fermi surface, the critical temperature of superconductivity is exponentially smaller than the characteristic phonon energy. Therefore such superconductors are found only at temperatures below a few Kelvin. Systems with flat energy bands have been suggested to cure the problem and provide a route to room-temperature superconductivity, but  previous studies are limited to only BCS models with an effective attractive interaction. Here we generalize Eliashberg's theory of strong-coupling superconductivity to systems with flat bands and relate the mean-field critical temperature to the microscopic parameters describing electron-phonon and electron-electron interaction. We also analyze the strong-coupling corrections to the BCS results, and construct the phase diagram exhibiting superconductivity and magnetic phases on an equal footing. Our results are especially relevant for novel quantum materials where electronic dispersion and interaction strength are controllable.
\end{abstract}

\maketitle

\section{Introduction}

The overarching idea in quantum materials is to design the electronic (or optical, magnetic, etc.) properties of materials to perform the desired functionality \cite{keimer17}. This goal is aided by generic models and concepts, such as specific lattice models that lead to certain topological phases.
Often the studied models and the resulting topological phases for electronic systems are noninteracting, and do not include the possibility of spontaneous symmetry breaking. However, such noninteracting models are platforms for exotic electron dispersions that provide a basis for studying symmetry broken interacting phases.
In particular, certain models support 
 approximate flat bands \cite{khodel1990superfluidity,heikkila2011flatbands,heikkilabook2016,tang2014strain,matsuura2013,peotta2015superfluidity,lothman2017universal,lieb1989,kauppila2016flat}, and here we consider microscopic mechanisms for symmetry breaking phases in such systems. 
 \begin{figure}
{\includegraphics[width=1\linewidth]{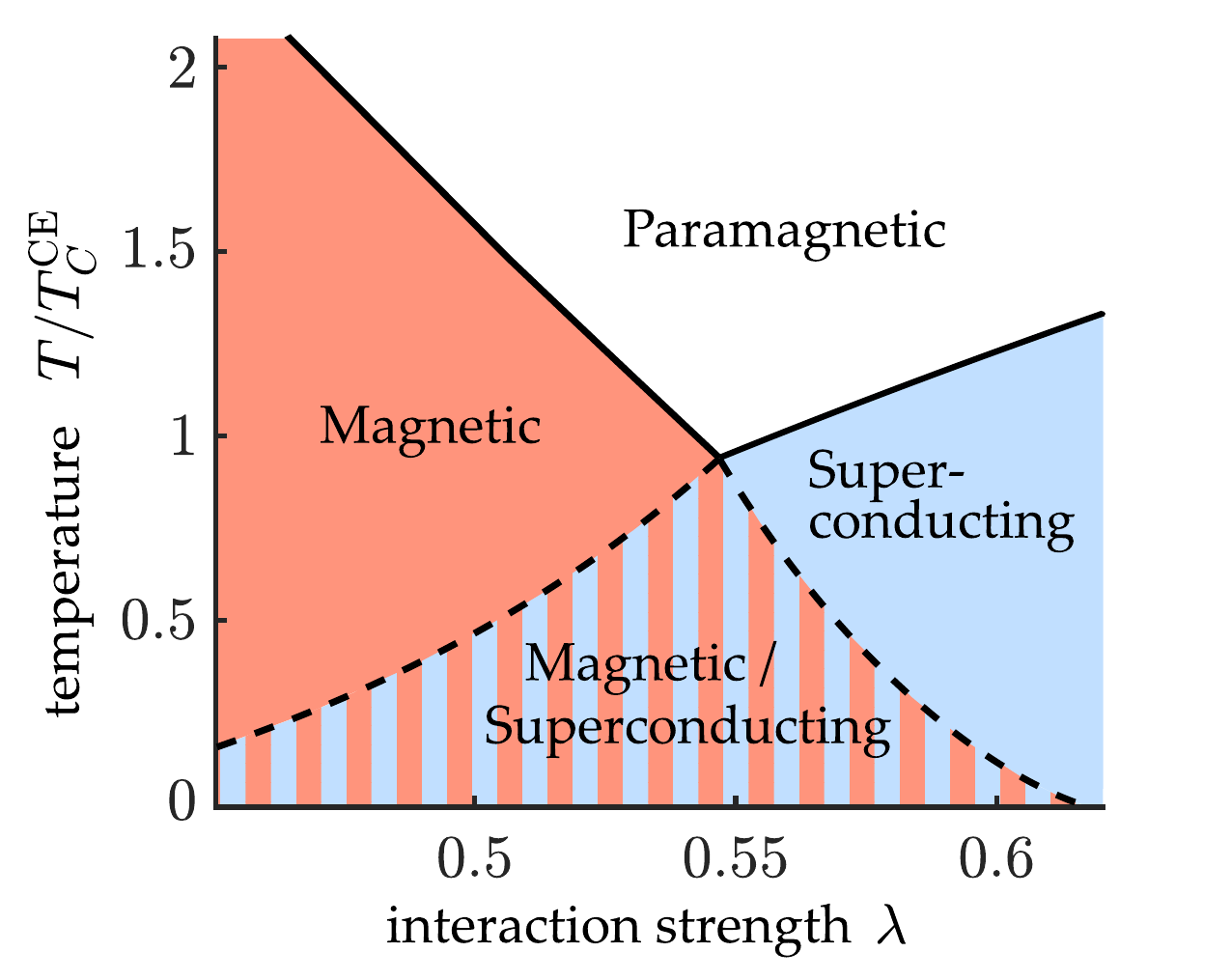}}
\caption{
Strong-coupling phase diagram for flat-band systems as a function of electron-phonon attraction $\lambda$ for electron-electron repulsion \(u=0.5\omega_E\) [Eq.(\ref{eq:intdef})].
\(T_C^{\rm CE}\) is the temperature at which the \(T_C\)'s of
magnetic and superconducting order coincide.
In the striped region these phases can form metastable domains inside the sample.
This diagram is for \(N\to\infty\). For finite \(N\) the overlap region between the phases is smaller.}
\label{fig:competition}
\end{figure}

We analyze the interplay of electron-phonon {\cite{frohlich50}} and (screened) electron-electron interaction in providing means for a symmetry-broken phase transition, thereby coupling together works on flat band superconductivity \cite{khodel1990superfluidity,kopnin2011high,peotta2015superfluidity,kauppila2016flat} with those on flat band (Stoner) magnetism \cite{lieb1989,tasaki1992ferromagnetism,mielke1993ferromagnetism,derzhko2007low,moon1995spontaneous,fertig1989energy}. 
In both cases the resulting mean-field critical temperature is linearly proportional to the coupling constant \cite{belyaev1961nature}, thus
 allowing for a very high critical temperature.
The two types of interaction mechanisms work in opposite directions, and in the case of weak interactions in a symmetric way. However, upon increasing the coupling strength the retarded nature of the electron-phonon interaction shows up --- as opposed to the instantaneous electron-electron interaction --- breaking the symmetry between the two. In particular, we generalize the Eliashberg's strong-coupling theory of superconductivity \cite{eliashberg1960interactions}, usually formulated for systems with a Fermi surface, for flat bands. As a result, we describe the dimensionless BCS attractive interaction \cite{bcs57} in terms of the electron-phonon coupling and the characteristic phonon frequency [Eq.~\eqref{eq:intdef}]. In addition, we provide the generalization of the well-known McMillan formula of strong-coupling superconductivity (for Fermi surface systems) \cite{mcmillan1968transition} to the case with flat bands
in Eq.~\eqref{eq:approximation}.

In addition to superconductivity,
we consider flat-band Stoner magnetism. Because of the retarded nature of the electron-phonon interaction, the combined interaction can simultaneously have attractive and repulsive components, and thus the system can be unstable with respect to both singlet superconductivity and magnetism (see a generic strong-coupling phase diagram in Fig.~\ref{fig:competition}). Often one of the phases still dominates and suppresses the other, but we find that when the critical temperatures of the phases are similar, both phases are local minima of the free energy at low temperatures. We find that their bulk coexistence and the resulting odd-frequency triplet superconducting order \cite{berezinskii1974new, matsumoto2012coexistence} 
is only realized as an unstable solution. On the other hand, these phases can form metastable domains inside the sample, and therefore odd-frequency triplet order parameter can appear at the domain walls.

The structure of this paper is as follows. In Sec.~\ref{Sec:Model} we introduce the model of surface bands with 
electron-phonon and Coulomb interactions.  
In Sec.~\ref{Eq:OrderedStates} we formulate the Eliashberg model extension for the surface bands, describe all possible ordered states that can appear within this model  and calculate the critical temperatures of the superconducting and antiferromagnetic  states. We study the competition and possible coexistence of these two types of ordering in Sec.~\ref{Sec:Competition}.
Conclusions are given in Sec.~\ref{Sec:Conclusions}.

\section{Model}
\label{Sec:Model}
As a low-energy model for the flat band, we assume two sublattices coupled through {an electronic Hamiltonian \cite{heikkila2011flatbands}
\begin{equation}
H_{{\rm el}, p} = 
\begin{pmatrix}
0& \varepsilon_p\\
\varepsilon_p &0
\end{pmatrix},\quad\text{with } \varepsilon_p = \varepsilon_0 \left(\frac{p}{p_{\rm FB}}\right)^N,
\label{eq:dispersion}
\end{equation}
}%
where an integer \(N\) parametrizes the flatness of the dispersion, and $\varepsilon_0$ is the energy at \(p=p_{FB}\). The model is electron-hole symmetric and the two energy bands have the dispersions \(\pm \varepsilon_p\). For large \(N\), the states with low momenta, \(|\vec p|\,{<}\,p_{FB}\), are almost at zero energy and the density of states is very high. %\TTH{This is otherwise ok, but in systems realizing the above dispersion, there are usually also other bands. Neglecting them amounts to setting an energy cutoff. Including them actually eventually would change the results, if $\omega_E$ is much larger than the energy of those bands. But it is probably ok not to mention it.} 
The states with momenta larger than \(p_{FB}\)
{do not contribute much to the momentum integrals due to {their} low density of states. {Therefore}, the results for large $N$ do not depend much on the momentum cutoff, as long as it is larger than \(p_{FB}\). In our model we take the cutoff to infinity and consider only the cases \(N>2\).} This is in contrast to models with isolated flat bands extending throughout the Brillouin zone.
The effects discussed below are mostly applicable also to such models (provided they have the type of sublattice degree of freedom discussed below) for large $N$, as long as $p_{FB}$ is taken as the size of the Brillouin zone. Equation \eqref{eq:dispersion} is approximately realized
for
the surface states of $N$-layer rhombohedrally stacked graphite. In that system the surface states delocalize into the  bulk at the edges of the flat band and this gives a momentum-dependent correction in the low-energy Hamiltonian \cite{kopnin2011high,Kauppila16}.
In the case of $N\to \infty$ the delocalization of the surface states to the bulk leads to strong amplitude mode fluctuations invalidating the mean-field theory \cite{Kauppila16}. Therefore, the theory considered in this paper is applicable to rhombohedral graphite only in the case where $N$ is not too large.

We model the electron-electron interaction as a repulsive on-site Hubbard interaction \cite{hubbard1963electron} with energy \(U\).
The magnitude
of $U$ depends on the microscopic details of the system and its environment.
The coupling between electrons and phonons, with strength \(g\), creates an effective attraction between the electrons and makes the system susceptible to superconductivity \cite{eliashberg1960interactions}.
We mostly consider Einstein phonons with constant energy \(\omega_q = \omega_E\) and discuss generalizations in the Supplementary Information.

The {total} Hamiltonian incorporating these effects is
\begin{equation}
\begin{split}
H &= \sum_{p,\sigma} \Psi_{p\sigma}^\dag
H_{{\rm el}, p}
\Psi_{p,\sigma} + \sum_{q,\rho} \omega_q b^\dag_{q,\rho} b^{\,}_{q,\rho} \\
&+ \frac{U}{2\mathcal{N}} \sum_{\substack{p,k,q\\\rho,\sigma,\sigma'}}
\psi_{p+q,\sigma\rho}^\dag \psi_{k-q,\sigma'\rho}^\dag \psi_{k,\sigma'\rho} \psi_{p,\sigma\rho} \\
&+ \frac{g}{\sqrt \mathcal{N}}\sum_{p,q,\sigma,\rho} (b^\dag_{-q,\rho} + b_{q,\rho}) \psi^\dag_{p+q,\sigma\rho} \psi_{p,\sigma\rho},
\end{split}\label{eq:hamiltonian}
\end{equation}
where \(\mathcal{N}\) is the number of lattice points in the system and \(\Psi_{p\sigma}^\dag = (\psi_{p\sigma A}^\dag, \psi_{p\sigma B}^\dag)\) is a pseudospinor in sublattice space. We assume that the low-energy states on the two sublattices \(\rho=A/B\) are spatially separated (e.g.~localized on the two surfaces in rhombohedral graphite), so that neither the electron-electron interactions nor the phonons couple them. The only coupling between the sublattices comes from the off-diagonal dispersion relation.
{In the Supplementary Information we also show that the flat band phenomenology applies to linear, graphene-like dispersion with an {electronic Hamiltonian
\begin{equation}
H_{{\rm el}, p} = \vF
\begin{pmatrix}
0 & p_x - i p_y \\ p_x + i p_y &0
\end{pmatrix}
,\tag{$1^\prime$}
\end{equation}
with} an energy cutoff \(\varepsilon_c\) and Fermi velocity \(v_F\), provided the interaction energy scales are large compared to \(\varepsilon_c\).} {Hence, our results may also apply as an effective model for twisted bilayer graphene close to its "magic" angle. \cite{cao2018unconvetional}}

In the theory of electron-phonon superconductivity of metals, the neglect of higher order diagrams in the perturbation theory is typically justified with the help of the Migdal theorem \cite{migdal1958interaction}. In that case, the expansion parameter gets an additional factor of \(\omega_E/E_F\), where \(E_F\) is the Fermi energy. Because of the Migdal theorem, the theory of superconductivity for metals is not strictly limited to weak coupling with respect to the interaction parameter.

In the flat band, however, the chemical potential is located at the bottom of the band and there is no Fermi energy with which to compare the Debye energy. Migdal's theorem cannot be used in this case. In the intermediate case of narrow electronic bands, corrections in the higher orders of the adiabatic parameter \(\omega_E/E_F\) have been studied in
Refs.~\onlinecite{grimaldi1995nonadiabatic,pietronero1995,grimaldi1995,botti2002nonadiabatic}
and the Eliashberg theory has been found also to be in agreement with Monte Carlo results in the weak coupling regime when \(\omega_E/E_F=1\) in Ref.~\onlinecite{esterlis2017breakdown}. We find that the diagrams beyond the mean-field approximation do not influence the self-energies significantly if 
the effective pairing constant introduced below in Eq.~\eqref{eq:intdef} is small \(\lambda\ll1\) and \(\omega_E,u\ll\varepsilon_0\). Moreover, although the mean-field theory is applied beyond its formal limits of validity in the strong-coupling regime,  this theory captures the interesting possibility that the retarded nature of the electron-phonon interaction can lead to the presence of attractive and repulsive components at the same time. As a result, the system can be simultaneously unstable with respect to the appearance of both singlet superconductivity and magnetism as discussed in Sec.~\ref{Sec:Competition}.

\section{Ordered states}
\label{Eq:OrderedStates}
The Hamiltonian \eqref{eq:hamiltonian} allows for a number of spontaneous symmetry breaking phases. We restrict our study to spatially homogeneous phases.
Therefore,  the order parameter can appear in the spin, sublattice (pseudospin) and electron-hole (Nambu) spaces. The general self-energy is
\begin{equation}
\Sigma(i\omega_n) = \sum_{i,j,k=0}^ 3 \Sigma_{ijk}(i\omega_n) \tau_i \sigma_j \rho_k,\label{eq:self-energy}
\end{equation}
where \(\tau_i\), \(\sigma_j\) and \(\rho_k\) are the Pauli matrices in electron-hole, spin and sublattice spaces, respectively. We characterize the different components \(\Sigma_{ijk}\) and determine their values within the self-consistent Hartree-Fock model. This reduces to solving a set of non-linear integral equations, known as Eliashberg equations in the context of conventional superconductors.

\begin{figure}
	 	\includegraphics[width=0.46\textwidth, trim={23mm, 183mm, 113mm, 17mm},clip]{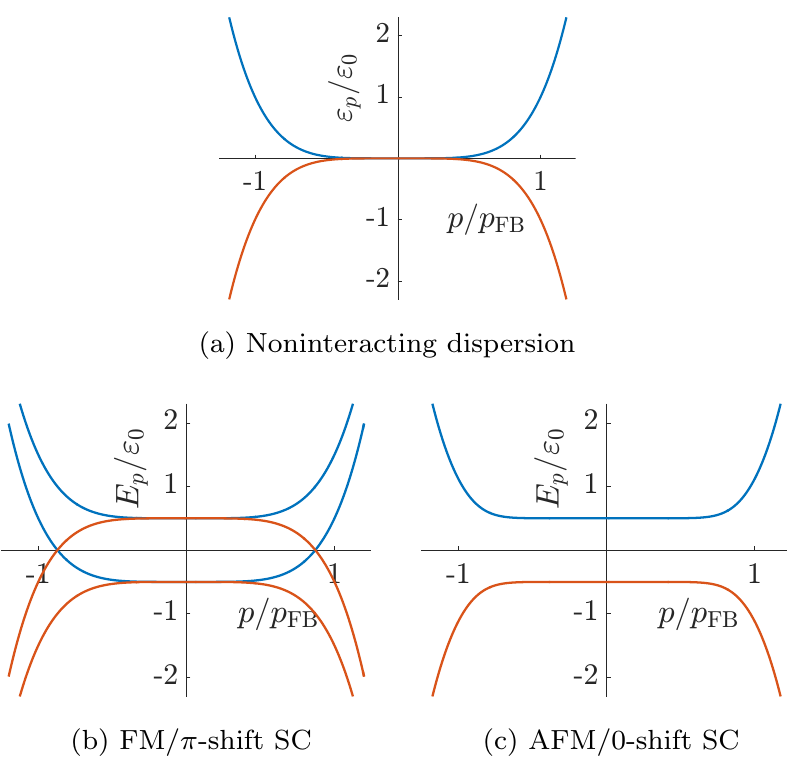}
  \caption{Quasiparticle dispersions $E(p)$ for different kinds of symmetry breakings with \(N=5\). a) In the
  non-interacting case,
  the spin-bands are degenerate with $E(p)=\pm \varepsilon(p)$. b) For the ferromagnetic (FM) or the superconducting (SC) phase with a \(\theta=\pi\) phase shift between the sublattices, one
 quasiparticle band is shifted up, and the other down in energy. In this case, no energy gap is opened. c) For the antiferromagnetic (AFM) or the SC phase with $\theta=0$ an energy gap is opened and quasiparticle bands are doubly degenerate.
 }
  \label{fig:dispersions}
\end{figure}

To explore the possible phases of the system, we first assume that the
\(U(1)\)-gauge symmetry
is broken, but the
\(SU(2)\)-spin rotation symmetry is not.
After fixing the overall phase of the superconducting order parameter, we are left with the self-energy $\Sigma_{000}(i\omega_n)$ and three degrees of freedom for the superconducting singlet order parameter: the magnitudes of the order parameter on the sublattices \(\Delta_A\) and \(\Delta_B\) and the relative phase \(\theta\).
{Choosing \(\theta=0\) leads to a gapped quasiparticle dispersion (Fig.~\ref{fig:dispersions}c), whereas \(\theta=\pi\) would imply a gapless dispersion (Fig.~\ref{fig:dispersions}b). Thus,} in the case of an instantaneous interaction the total energy is minimized when \(\theta=0\) and $\Delta_A=\Delta_B$. 
Generalizing the above to the frequency dependent interactions, we choose the singlet to be proportional to the \(\tau_2 \sigma_2 \rho_0\)-component, whose magnitude and the functional form are obtained from the self-consistency equation. The self-energy for the fermionic Matsubara frequency \(\omega_n\) is
\begin{equation}
\Sigma_{\rm SC}(i\omega_n) = -i\Sigma^\omega_n \mathbbm{1} + \phi_n \tau_2 \sigma_2,\label{eq:scse}
\end{equation}
where \(\Sigma_n^\omega = (1-Z_n)\omega_n\) is the frequency renormalization by the retarded interaction \cite{eliashberg1960interactions}. To simplify the equations, we define renormalized frequencies \(\tilde\omega_n=Z_n\omega_n\). We use the symbol \(\phi_n\) for the 'bare' singlet order parameter and \(\Delta\) for the maximum value of the renormalized singlet order parameter \(\Delta_n \equiv \phi_n/Z_n\)
related to the energy gap.

When $SU(2)$-spin rotation symmetry is broken but
{$U(1)$-gauge symmetry} is not,
{the self-energies describe
the frequency renormalization}
and
the magnetization.
After fixing the direction of the magnetization on one sublattice, the relevant degrees of freedom are reduced to three {similarly} as in the superconducting case. These can be chosen as the magnitudes of the magnetizations
in the two sublattices \(h_A\), \(h_B\) and the relative angle $\varphi$
between their directions.
The quasiparticle dispersion
in the magnetic case is
the same as in the superconducting case
if we identify \(\Delta_{A,B}=h_{A,B}\) and \(\theta = \pi{-}\varphi\) (see Fig.~\ref{fig:dispersions}).
{In this case, the relative angle \(\varphi=0\) leads to a gapless quasiparticle dispersion (Fig.~\ref{fig:dispersions}b), and \(\varphi=\pi\) to a gapped dispersion (Fig.~\ref{fig:dispersions}c).} {Thus, the} energy minimum is obtained with \(h_A = h_B\) and
\(\varphi=\pi\). The stable magnetization is hence antiferromagnetic, with opposite magnetizations on the two sublattices,
so that the self-energy is
\begin{equation}
\Sigma_{\rm AFM}(i\omega_n) = -i\Sigma^\omega_n \mathbbm{1} + h_n \tau_3 \sigma_3 \rho_3,\label{eq:afmse}
\end{equation}
where \(h_n\) is the frequency dependent exchange field. This result agrees with DFT studies on rhombohedral graphite \cite{pamuk2017magnetic}
and similar magnetization structure has been predicted also in the case of flat bands appearing at the zig-zag edges of graphene nanoribbons \cite{fujita1996peculiar, fernandez2008prediction, son2006half}. We also note that the AFM state is insulating (see Fig.~\ref{fig:dispersions}c). If the non-interacting dispersion is completely flat at zero energy, the sublattices are uncoupled and the antiferromagnetic state is degenerate with the ferromagnetic \(\varphi=0\) state. 

By calculating the Hartree-Fock self-energies, we find the {self-consistency equations, from which we can determine the values of the self-energy terms}. For {the} superconducting (SC) self-energy \eqref{eq:scse}, they are 
\begin{align}
\phi_n &= 2 T \!\!\sum_{m=-\infty}^\infty\!\! ( \lambda_{nm} {-} u ) \int_0^\infty \frac{\ddp p}{p_{FB}^2} \frac{\phi_m}{\tilde\omega_m^2 + \varepsilon_p^2 + \phi^2_m},\label{eq:eliashberg_sc_phi}\\
Z_n &= 1 + 2 T\!\!\sum_{m=-\infty}^\infty\!\! \lambda_{nm} \frac{\omega_m}{\omega_n}
\int_0^\infty \frac{\ddp p}{p_{FB}^2} \frac{Z_m}
{\tilde\omega_m^2 + \varepsilon_p^2 +\phi_m^2},\label{eq:eliashberg_sc_Z}
\end{align}
where the interaction kernel is \(\lambda_{nm} = \lambda\omega_E^3/\left[\omega_E^2 + (\omega_n-\omega_m)^2\right]\). The functional form of the interaction kernel is determined by the phonon propagator from which it is derived. The width in frequency space is determined by the characteristic phonon frequency, which in this case is the Einstein frequency \(\omega_E\). The effective interaction constants in the flat band are 
\begin{align}
\lambda = \frac{g^2}{\omega_E^2} \frac{\Omega_{\rm FB}}{ \Omega_{\rm BZ}},\quad 
u = \frac{U \Omega_{\rm FB}}{\Omega_{\rm BZ}}\label{eq:intdef},
\end{align}
where \(\Omega_{\rm FB}\) and \(\Omega_{\rm BZ}\) are the momentum-space areas of the flat band and of the first Brillouin zone, respectively.

For an antiferromagnet (AFM)
with self-energy \eqref{eq:afmse}, the self-consistency equations are
\begin{align}
h_n &= 2 T \!\!\sum_{m=-\infty}^\infty\!\! ( u {-} \lambda_{nm} ) \int_0^\infty \frac{\ddp p}{p_{FB}^2} \frac{h_m}{\tilde\omega_m^2 + \varepsilon_p^2 + h_m^2},\label{eq:eliashberg_h_phi}\\
Z_n &= 1 + 2 T \!\!\sum_{m=-\infty}^\infty\!\! \lambda_{nm} \frac{\omega_m}{\omega_n}
\int_0^\infty \frac{\ddp p}{p_{FB}^2} \frac{Z_m}
{\tilde\omega_m^2 + \varepsilon_p^2 +h_m^2}.\label{eq:eliashberg_h_Z}
\end{align}
Superconductivity and magnetism are thus symmetric with each other also on the level of the self-consistency equations, but with the roles of \(u\) and \(\lambda_{nm}\) switched. {Tovmasyan et al. have shown that this duality is also broken by taking into account higher order terms in the perturbation theory \cite{tovmasyan2016effective}.}

\begin{figure}
\includegraphics[width=0.465\textwidth, trim={20mm, 167mm, 111mm, 17mm},clip]{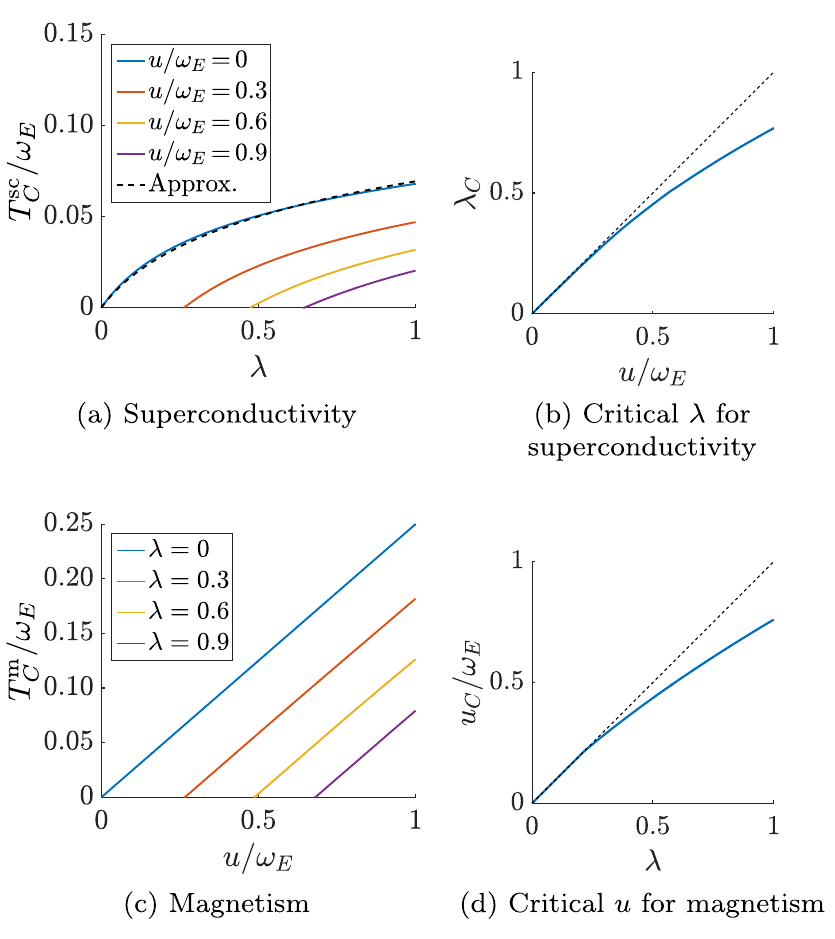}
  \caption{Critical temperatures for superconducting and magnetic phases for \(N\to\infty\). (a) Superconductivity is suppressed when \(\lambda \lesssim u/\omega_E\). Above the critical point \(\lambda_C(u)\), \(\TCsc\) is linear in \(\lambda\). With increasing \(\lambda\), the electron-phonon renormalization increases and this limits the critical temperature. Dashed line is the approximation in Eq.~\eqref{eq:approximation}. (b) Critical interaction strength for superconductivity as a function of u. When \(\lambda<\lambda_C(u)\), superconductivity is suppressed. Dashed line is the instantaneous approximation. (c) Magnetism is suppressed when \(u/\omega_E \lesssim \lambda\). Above the critical point \(u_C(\lambda)\), \(\TCm\) is linear in \(u\). (d) Critical interaction strength for magnetism as a function of electron-phonon interaction. When \(u<u_C(\lambda)\), magnetism is suppressed. Dashed line is the instantaneous approximation. {In this figure, we do not take into account the possible magnetic instability of the superconducting state, or vice versa.}} \label{fig:critical_interactions}
\end{figure}

{To solve the self-consistency equations (\ref{eq:eliashberg_sc_phi}--\ref{eq:eliashberg_h_Z}), we truncate the Matsubara sums with a cutoff \(\omega_C \sim 10\omega_E\). This causes no numerical error if we use the pseudopotential trick and simultaneously replace \(u\) with an effective value \(u^*\), which depends on the cutoff \cite{morel1962calculation}. For superconductivity (magnetism) cutting off high-energy scatterings is compensated by a reduction (increase) in the low-energy effective interaction.

After the pseudopotential trick, the solutions are found by a fixed-point iteration. The iteration is continued until all of the components have converged. The fixed-point method only finds the stable solutions, to find the unstable solutions we used a solver based on Newton's method.}

The number of parameters in Eqs.~(\ref{eq:eliashberg_sc_phi}--\ref{eq:eliashberg_h_Z}) can be reduced by defining new interaction constants \(\tilde\lambda \equiv \lambda (\omega_E/\varepsilon_0)^{2/N}\) and \(\tilde u = u\omega_E^{2/N-1}/\varepsilon^{2/N}\), so that one parameter is eliminated completely and the results become proportional to \(\omega_E\).

For weak coupling \(\lambda \ll 1\) the frequency dependence of \(\lambda_{nm}\) can be disregarded and we can approximate \(Z \approx 1\) and \(\Delta \approx \phi\). Assuming \(\lambda\omega_E>u\), the superconducting gap at \(T=0\) and the critical temperature are 
\begin{align}
\frac{\Delta_0}{\omega_E} &= \frac{1}{2}\left[\frac{(\tilde\lambda-\tilde u) \sqrt{\pi} \Gamma(\frac{1}{2}{-}\frac{1}{N})}{N \sin(\frac{\pi}{N} )\Gamma(1{-}\frac{1}{N})} \right]^{\frac{N}{N{-}2}},\label{eq:Delta0}\\
\frac{\TCsc}{\omega_E} &= \frac{1}{2\pi}\left[\frac{(\tilde\lambda-\tilde u) \zeta(2{-}\frac{2}{N})\left(2^{2{-}\frac{2}{N}}{-}1\right)}{N \sin(\frac{\pi}{N})} \right]^{\frac{N}{N-2}}.\label{eq:instant_TC}
\end{align}
{These results are valid for \(N>2\) as the momentum integrals diverge without a cutoff for \(N\leq2\)}. {Note that the $T=0$ limit can thus be taken before the flat-band limit of large $N$.} Analogous results have been obtained before within the BCS model in Ref.~\onlinecite{kopnin2011high}. For large \(N\), \(\Delta_0\) is linear in the coupling and its magnitude is proportional to the phonon energy scale. Hence the associated critical temperature can be very large. Relabeling \(\Delta_0 \rightarrow h_0\) and \(\tilde\lambda\leftrightarrow \tilde u\), we find similar equations for magnetism.
Here \(h_0\) is the magnetic order parameter at \(T=0\).

At strong coupling, the retardation matters and the results for magnetism and superconductivity diverge from each other. For superconductivity, we can improve on the weak coupling result by including some of the corrections from the Eliashberg theory when $N {\to} \infty$. We still neglect the full frequency dependence, but include the electron mass renormalization as a static factor \(Z_0 = 1{+}\lambda\). The order parameter at zero temperature becomes
\begin{equation}\label{eq:Delta0FB}
\Delta_0 = \frac{\lambda\omega_E{-}u}{2(1+2\lambda)}.
\end{equation}
In metals with a Fermi surface\cite{carbotte1990properties}, the electron-phonon interaction renormalizes the pairing potential with the factor of \(1{+}\lambda\) instead of \(1{+}2\lambda\) as in Eq.~\eqref{eq:Delta0FB}. Thus, for weak coupling, the electron-phonon renormalization is more effective in the flat band than in the usual metals. This difference is more pronounced at strong coupling, as we see next.

By linearizing Eqs.~\eqref{eq:eliashberg_sc_phi} and \eqref{eq:eliashberg_sc_Z} with respect to \(\phi\), we can solve for the critical temperature, see Fig.~\ref{fig:critical_interactions}a.
We find that when \(N\to\infty\), the critical temperature scales as \(\TCsc \propto \lambda^{0.2}\omega_E\) for large \(\lambda\). In metals\cite{carbotte1990properties} the asymptotic scaling goes as \(\TCsc\propto\lambda^{1/2}\omega_E\). 

When \(u\neq0\), there is a critical point \(\lambda_C\) such that for \(\lambda<\lambda_C\) there is no superconducting transition at any temperature. For small \(u/\omega_E\), \(\lambda_C\) is linearly proportional to the Coulomb interaction. For large \(u\), \(\lambda_C\) increases sublinearly, see Fig.~\ref{fig:critical_interactions}b.

An approximate numerical equation for \(\TCsc\) is
\begin{equation}
\TCsc = \frac{\lambda\omega_E-u(1-0.3u/\omega_E)}{4(1+2.6\lambda^{0.8})}.\label{eq:approximation}
\end{equation}
This is a flat band analogue of the McMillan equation \cite{mcmillan1968transition} which for the the conventional superconductors incorporates the Eliashberg and Coulomb corrections to \(\TCsc\). {The \(u^2\)-term in the numerator accounts for the retardation correction to \(\lambda_C\) as in Fig.~\ref{fig:critical_interactions}b. The form of the denominator is chosen to show the \(\lambda^{0.2}\) power law behaviour for large \(\lambda\). The factor \(2.6\) is obtained by a fit in the region \(\lambda<1\) for \(u=0\). The fit is shown as the dashed line in Fig.~\ref{fig:critical_interactions}a.}

\begin{figure}[b]
\includegraphics[width=0.85\linewidth, trim={0mm 0mm 0mm 0mm},clip]{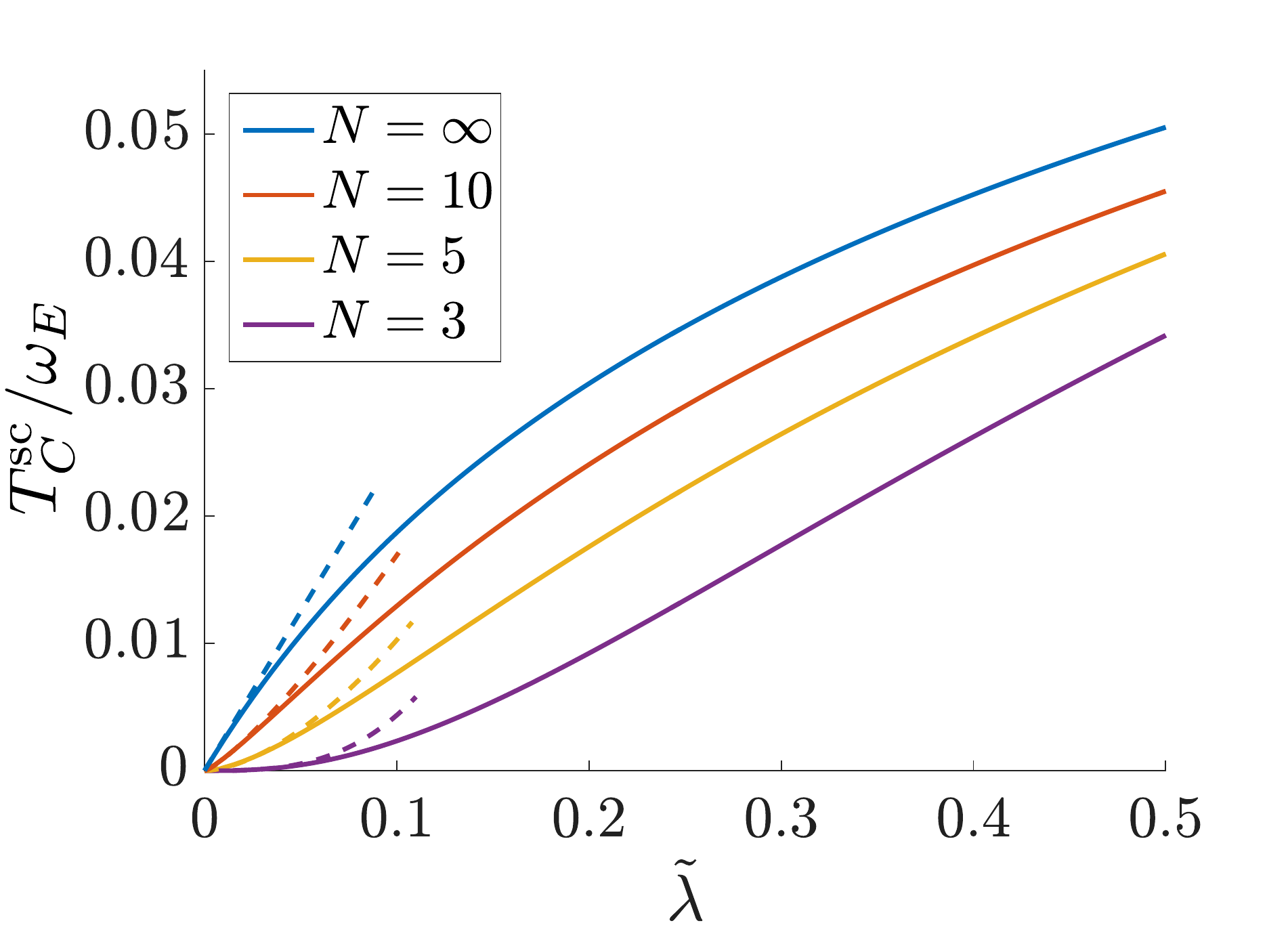}
\caption{Effect of finite \(N\) on critical temperature when \(u=0\). For small \(\tilde\lambda\), the results coincide with the instantaneous approximation of Eq.~\eqref{eq:instant_TC} (shown with the dashed lines). For large \(\tilde\lambda\), the electron-phonon renormalization limits the increase in \(\TCsc\).}\label{fig:finiteN}
\end{figure}

The ratio \(\Delta_0/\TCsc\) is not constant, but depends on both \(N\) and \(\lambda\). For \(N\to\infty\), the ratio has the value 2 for weak coupling and increases as \(\lambda\) increases. For \(\lambda {=} 1\) the ratio is 2.56. For the critical temperature at finite \(N\), see Fig.~\ref{fig:finiteN}. 

The phenomenology of the magnetism can be understood as follows.
According to the Stoner criterion, the magnetization is related to the competition between the exchange energy gain and the kinetic energy penalty from moving electrons from one spin band to another. For a flat band with \(N\to\infty\), there is no kinetic energy penalty, and at zero temperature with \(\lambda\,{=}\,0\) even a small exchange interaction leads to a complete magnetization of the flat band. In the presence of the electron-phonon interaction the competition is between the exchange energy gain and the electron-phonon energy penalty
which coincide at \(u=u_C\).
If we can neglect the retardation, the total interaction in Eq.~\eqref{eq:eliashberg_h_phi} is \(u-\lambda\omega_E\). The flat band is completely magnetized when \(u > u_C \approx \lambda\omega_E\). Due to retardation, for large \(\lambda\) the critical point is reduced from the linear estimate, see Fig.~\ref{fig:critical_interactions}d.

Above, we have discussed the superconducting order parameter \(\phi\). The other important property of the superconducting state is the existence of a supercurrent. In the flat band the electronic group velocity vanishes and it is not immediately clear that there can be a finite supercurrent. However, the flat band surface states of superconducting rhombohedral graphite do support a finite supercurrent \cite{kopnin2011surface} and similarly it is known that quantum hall pseudospin ferromagnets can support a finite pseudospin supercurrent \cite{moon1995spontaneous}. More generally, Peotta and T\"orm\"a \cite{peotta2015superfluidity} have shown that for a topological flat band there is an additional {geometric} contribution to the superfluid weight so that the critical current is finite.
As we have not fixed the underlying topology in our model, it can be applied to topologically non-trivial flat bands.

\begin{figure}
{\includegraphics[width=0.75\linewidth]{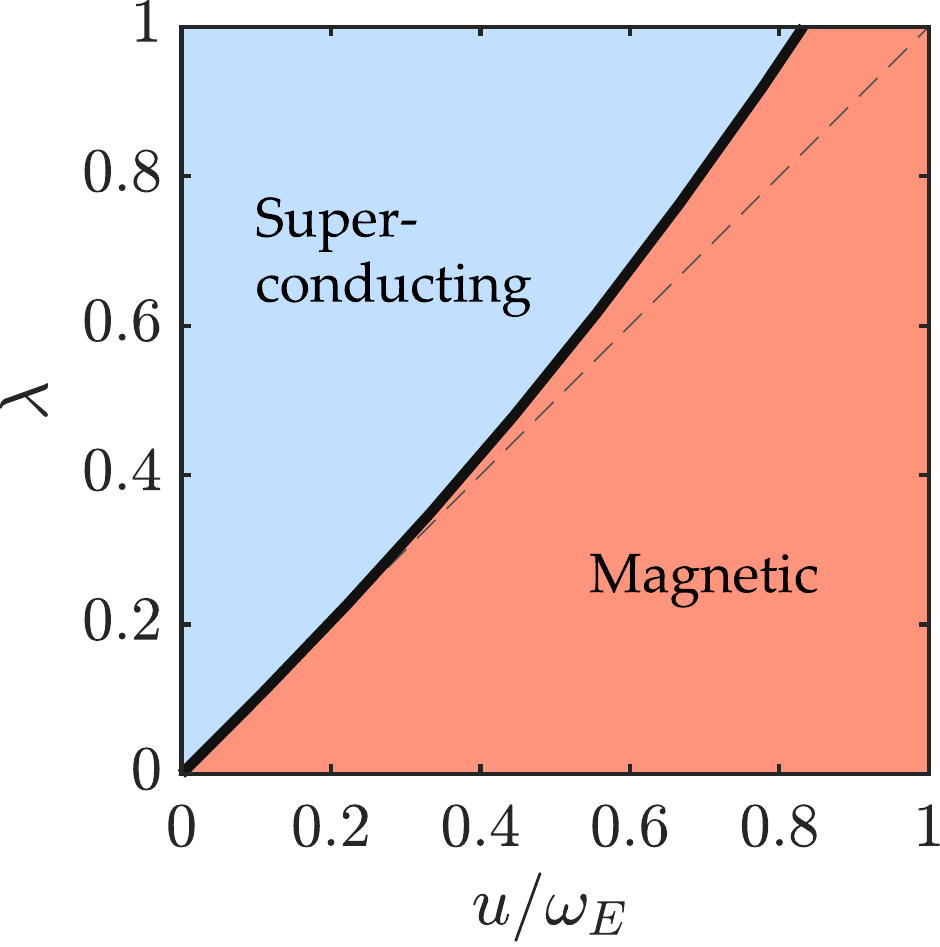}}
\caption{Mean-field phase diagram for \(N=\infty\) obtained by solving the curve on which the critical temperatures for superconductivity and antiferromagnetism are equal. Thin dashed line shows the phase boundary \(\lambda=u/\omega_E\) in the case of instantaneous interactions. When the energy scales of interactions are small compared to \(\omega_E\) we recover the BCS results.
{The phase diagram for finite \(N\) looks similar but the retardation effects are weaker, so that the deviation from the BCS approximation is smaller.}}
\label{fig:phasediagram}
\end{figure}

As one can see the Eliashberg model describes the nucleation of both the magnetic and superconducting phases which 
can have rather close critical temperatures as shown in Fig.~\ref{fig:critical_interactions}. 
In the next section we consider the non-linear problem by calculating the entire phase diagram of the ordered states
to study the competition and the possible coexistence between the superconductivity and antiferromagnetism.  

\section{Competition between the phases.} 
\label{Sec:Competition}
If the electron-phonon interaction is approximated as instantaneous, we can sum the two interactions together and have either a total interaction which makes the normal state unstable to the superconducting transition (\(\lambda\omega_E{-}u>0\)) or to the magnetic transition (\(\lambda\omega_E{-}u<0\)), but not to both at once. {On the other hand,}
if the electron-phonon interaction is retarded, the situation is different as the total interaction {can} be attractive for low frequencies, but repulsive for high frequencies. There is then a parameter range in which both phases are
{local minima of the free energy}. This occurs when \(\lambda\) is large enough to overcome the suppressing effect of \(u\) in the case of superconductivity ({\(\lambda > \lambda_C(u)\) in} Fig.~\ref{fig:critical_interactions}b), but at the same time \(u\) is large enough to overcome {the suppressing effect of} \(\lambda\) and create a magnetic instability ({\(u>u_C(\lambda)\) in} Fig.~\ref{fig:critical_interactions}d).

We determine the phase diagram {of the system}
by calculating the phase with a higher critical temperature {as a function of $u$ and $\lambda$} (Fig.~\ref{fig:phasediagram}). The phase diagram is almost symmetric with respect SC and AFM phases
{except that} the lack of retardation in electron-electron repulsion favors the AFM phase for strong coupling.

Even if there is a parameter region in \(T\), \(u\) and \(\lambda\) where both SC and AFM self-consistency equations have a finite solution, it does not mean that both phases are necessarily simultaneously present. To determine the stability, we construct the coupled self-consistency equations in the case when both {order parameters} are non-zero and interact with each other \cite{supplementary}. By linearizing the coupled self-consistency equation with respect to SC, and solving the AFM part fully, the stability of the AFM phase with respect to SC transition can be determined, and vice versa. Fig.~\ref{fig:competition} shows the region in \(\lambda\)--\(T\) space with fixed \(u\), where the two phases are stable. The figure shows that in the region where SC is dominant, the AFM phase is unstable near the expected second order transition (the solid line {between the magnetic and paramagnetic phases}), but becomes a local minimum of free energy at lower temperatures. {The same happens for superconductivity when the AFM phase dominates.}
The transition between SC and AFM phases is of the first order.

When discussing superconductivity in the presence of an exchange field (either induced or spontaneous), we have an additional ingredient in the self-energy, namely the superconducting triplet order parameter \cite{berezinskii1974new,balatsky1992new}, which has been discussed in the context of the Eliashberg model in Ref.~\onlinecite{kusunose2012strong}. The triplet is spatially isotropic, and in order to satisfy the fermionic antisymmetry, it has to be odd in frequency. It is generated in the self-energy only when there is an odd-frequency component in the interaction. In the retarded interaction, this is always satisfied. When calculating the stability of AFM with respect to SC, the triplet appears in the linear order. It hence modifies the boundaries of the region where both AFM and SC are stable. We have taken this effect into account in Fig.~\ref{fig:competition}.

Besides the competition between AFM and SC phases, {we need to consider the possibility of} a coexistence phase
in the dashed region of Fig.~\ref{fig:competition}, where both phases can show up alone. We indeed have numerically found such a coexistence solution, but tests based on fixed-point iteration revealed it to be unstable at every temperature that we checked. This finding is in accordance with a
{simplified} model where
both  interaction channels are instantaneous
{and independent of each other \cite{supplementary}.}

\begin{figure}
{\includegraphics[width=0.8\linewidth]{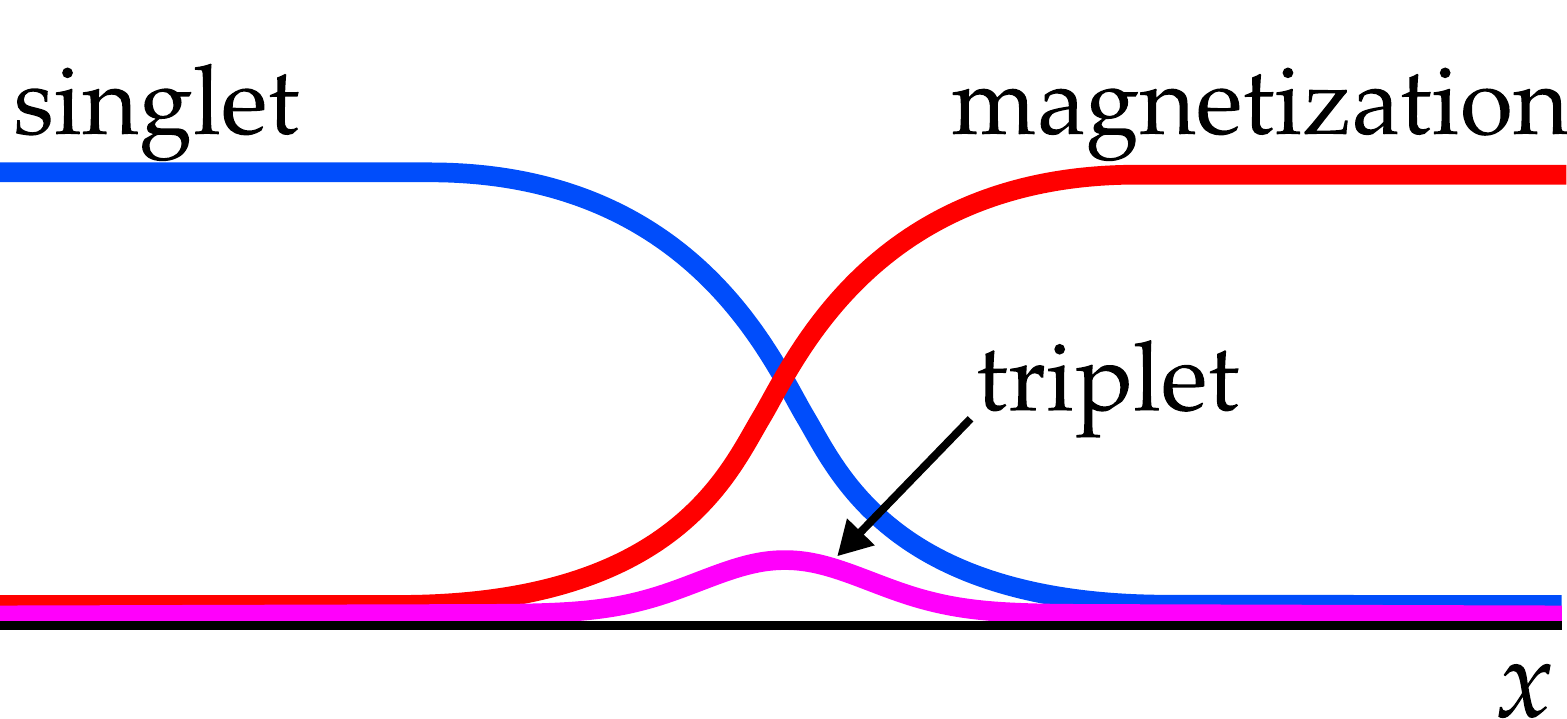}}
\caption{Sketch of a domain wall between magnetic (red) and superconducting (blue) domains. At the domain wall a triplet component (purple) is induced.}
\label{fig:domain_wall}
\end{figure}

However, the fact that the two phases
{are simultaneously local minima of the free energy} suggests that this system could have domains of antiferromagnetic order coexisting with superconducting domains. Such domains would be separated by a domain wall mixing the two kind of phases and inducing odd-frequency triplet pairing, as
{schematically illustrated} in Fig.~\ref{fig:domain_wall}. {In addition to providing a mechanism for the appearance of odd-frequency triplet pairing, the domain walls can support interesting excitations. In particular, it is known that flat band ferromagnets can support interesting topological and domain wall excitations in the form of different kinds of spin textures \cite{moon1995spontaneous, Pikulin2016}, and various combinations of spin textures and superconductivity may lead to the appearance of Majorana zero modes \cite{Choy11, Pascal13, Klinovaja13,Franz13,pientka2013topological}. Also, alternatively to the intrinsic domain structure generation, the ferromagnetic superconductors can support different types of nonuniform magnetic order and spontaneous vortex states \cite{Anderson59, Bulaevskii85,Vinnikov17}. A detailed analysis of different possibilities goes beyond the scope of this paper.}

\section{Conclusions}
\label{Sec:Conclusions}
We have proposed a simplified model of a flat band system with a retarded electron-phonon interaction and a repulsive Hubbard interaction. For this model, we have determined the self-consistency equations in the Hartree-Fock approximation and all the possible homogeneous phases. Antiferromagnetism and superconductivity are essentially symmetric in this system, with the only difference coming from the retardation of the electron-phonon interaction. For large \(\lambda\), the retardation suppresses the increase in \(\Delta\) more effectively in a flat band than in metals with a Fermi surface. We find that the retardation also creates a situation in which both phases are separately local minima of the free energy suggesting a possibility of coexisting antiferromagnetic and superconducting domains inside the sample.

Our results indicate how flat-band superconductivity can be generated from electron-phonon interaction, and provides means to estimate the mean-field critical temperature when the details of the electron-phonon coupling and the screened interaction are known. The superfluid transition in low-dimensional systems occurs in the form of a Berezinskii-Kosterlitz-Thouless (BKT) transition at a temperature that is lower than the mean-field transition temperature. That the latter is non-zero is ensured by the possibility of having a non-vanishing supercurrent (see, for example, Refs.~\onlinecite{kopnin2011surface,peotta2015superfluidity,kauppila2016flat}) in a flat-band superconductor.
Our results are of relevance in designing novel types of quantum materials for the interplay of superconducting and magnetic order, and the search of systems exhibiting exotic superconductivity with a very high critical temperature, up to room temperature. They may also shed light on recent evidence of
high-temperature superconductivity in graphite interfaces \cite{precker2016}.

{Our results could also explain some of the phenomena associated with the recent experiments on bilayer graphene \cite{cao2018correlated,cao2018unconvetional}. ({For a more microscopic description of that case within the BCS model, see Refs.~\onlinecite{peltonen2018mean,wu2018theory}.)} In the experiment, the twist angle between two superimposed graphene layers is chosen to a certain magic angle, so that the two Dirac cones in the graphene layers hybridize, forming a pair of flat bands. {Our model can be adjusted to describe this situation with small changes, see the Supplementary Information for details.} When the chemical potential was tuned to lower of these bands, the system became an insulator. From our point of view, this could be the insulating AFM state we describe. When the chemical potential is tuned slightly off from the flat band, a superconducting dome in \(T-\mu\) phase diagram was observed on both sides. These domes can be the s-wave SC phases we describe here. The competition between the particle-hole (AFM) and the particle-particle (SC) channels in the presence of the chemical potential has been considered by L\"othman and Black-Schaffer in Ref.~\onlinecite{lothman2017universal}, and for a range of parameters, they reproduce a similar phase diagram near the flat band, with the AFM state at the level of the flat band and two superconducting domes with doping away from the flat band (see Fig 2b in the paper). 
In the experiments, SC domes are only observed on the hole-doped side. The electron-doped side exhibits only insulating behavior near the flat band. One possible explanation is the difference in screening, which changes the relative magnitude of repulsive and the attractive interactions, so that the AFM state covers the SC domes completely. {However, we leave the detailed treatment of the effects of doping and screening (both intrinsic and that provided by the environment) for further work.}}

\section*{Acknowledgements}

We thank Sebastiano Peotta, Long Liang and P\"aivi T\"orm\"a for helpful comments.
This project was supported by the
Academy of Finland Key Funding (Project No. 305256),
Center of Excellence (Project No. 284594) and
Research Fellow (Project No. 297439)
programs.

\bibliography{refs}

\clearpage
\appendix
\counterwithout{equation}{section}
\setcounter{equation}{0}
\renewcommand{\theequation}{S\arabic{equation}}
\widetext

\section{Note on interaction constants and the tight binding model}

The Hamiltonian studied in the main text is derived from a
tight binding model, in which the size of the system is naturally
characterized by the number of lattice sites \(N\). However, in
approximating the sum over all momenta, it is more natural to use the
area $A$ of the system. The ratio \(A/N\) is the area of the real
space unit cell \(A_c\), which in turn is inversely proportional to
the area of the first Brillouin zone \(\Omega_{\rm BZ}\). For an infinite system, the momentum sum can be written in the following form.
\begin{equation}
\frac{1}{N} \sum_p = \frac{A}{N} \int_{\rm BZ} \frac{\dd[2]{\vec p}}{(2\pi)^2} = \frac{A_c \pi p_{\rm FB}^2}{2\pi^2} \int_0^{p_c} \frac{\dd{p} p}{p_{\rm FB}^2} = 2\frac{\Omega_{\rm FB}}{\Omega_{\rm BZ}} \int_0^{p_c} \frac{\dd{p} p}{p_{\rm FB}^2},
\end{equation}
where \(\Omega_{\rm FB} \equiv \pi p_{\rm FB}^2\) is the area of the flat band {and \(p_c\) is the momentum cutoff}. We define a shorthand for the sum/integral over momenta and Matsubara frequencies
\begin{equation}
\sum_{\vec p, m} = 2 T\sum_{\omega_m} \int_0^{p_c} \frac{\dd{p} p}{p_{\rm FB}^2}.
\end{equation}

We find that the effective interactions on the flat band are characterized by the constants
\begin{align}
u &\equiv \frac{U \Omega_{\rm FB}}{\Omega_{\rm BZ}},\\
\lambda_{nm} &= -\frac{g^2}{2} \frac{\Omega_{\rm FB}}{\Omega_{\rm BZ}} D(i\omega_m {-} i\omega_n) = \frac{\lambda \omega_E^3}{\omega_E^2 + (\omega_m {-} \omega_n)^2}, \qq{with} \lambda \equiv \frac{g^2}{\omega_E^2} \frac{\Omega_{\rm FB}}{\Omega_{\rm BZ}}.
\end{align}
where \(D(z) = -2\omega_E/(\omega_E^2 - z^2)\) is the phonon
propagator. In other words, interactions are proportional to the ratio
between the area of the flat band and that of the first Brillouin zone.% For rhombohedral graphite the ratio is \(0.03^2\approx 10^{-3}\).

\section{Self-energy components}

In total, there are \(4^3 = 64\) combinations of Pauli matrices in
spin, Nambu and sublattice spaces. The ones off-diagonal in sublattice
space are not possible (see below), as the interactions in the model
do not couple the two sublattices. This reduces the number by a factor of 2. We are left with 16 components symmetric and 16 components antisymmetric in sublattice index \(\rho\). Of these 32 components, the 16 components diagonal in Nambu space are associated with non-superconducting properties. The \(\tau_0 \sigma_0 \rho_0\)-component renormalizes the frequencies in the propagator \cite{carbotte1990properties}. It vanishes if the interaction is instantaneous and is always present if the interaction has a nontrivial frequency structure. The \(\tau_3 \sigma_0 \rho_0\)-component on the other hand renormalizes the chemical potential, and is usually induced by finite temperature or interaction effects. In this model it vanishes because of the electron-hole symmetry of the model at half-filling. There could in principle also be a term proportional to \(\tau_0 \sigma_0 \rho_3\). Its effect would be to renormalize frequencies antisymmetrically in the sublattices. However, it is not induced by any of the other terms, so the only way to have it would be by spontaneous symmetry breaking. It should be odd in frequency, and this makes it vanish in the BCS limit. Even in Eliashberg theory, it is unlikely, as it is supported only by the odd-frequency part of the interaction. In this text, we do not consider this and other antisymmetric frequency renormalization components any further.

The remaining 12 components diagonal in Nambu space are due to magnetism. The magnetization direction on one sublattice is described with the three components \(\sigma_{i}\tau_3\). We can parametrize the six degrees of freedom associated with magnetization with the overall magnitude and the relative magnitude of the order parameter, two angles for the overall magnetization direction, and two angles for the relative direction. The six other terms of the form \(\sigma_i \tau_0 \rho_j\) are spin-antisymmetric frequency renormalization components.

The 16 components off-diagonal in Nambu space are associated with
superconductivity. Four of these are associated with the singlet and
its phase and the sublattice: overall phase, relative phase between
the sublattices, overall magnitude of the order parameter and the
relative magnitude between the sublattices. The remaining 12 off-diagonal Nambu components describe the three components of the triplet and its phase and sublattice degrees of freedom. Because both the spatial and the spin parts of the triplet are symmetric, it must be odd in frequency to preserve the fermionic antisymmetry \cite{berezinskii1974new}. Such components are supported by the odd-frequency part of the electron-phonon interaction \cite{matsumoto2012coexistence}.

\section{Hartree-Fock self-energies}

Starting from the Hamiltonian and using the above definitions for the interactions, we can write the self-energies in the Hartree-Fock approximation as
\begin{align}
\Sigma_{\rm H}^{\rm c} &= - u \sum_{\vec p, m,\rho} P_\rho \Tr[ P_\rho G(\vec p, i\omega_m) ]\\
\Sigma_{\rm F}^{\rm c} &= u \sum_{\vec p, m,\rho} P_\rho G(\vec p, i\omega_m) P_\rho\\
\Sigma_{\rm H}^{\rm ph} &= 0\\[0.7em]
\Sigma_{\rm F}^{\rm ph}(i\omega_n) &= - \sum_{\vec p, m,\rho} \lambda_{nm} P_\rho G(\vec p, i\omega_m) P_\rho,
\end{align}
where \(P_{\rho}\) is the projection operator to sublattice $\rho$,
\(\rho\in A,B\). It is immediately clear from the above expressions
that the self-energy cannot have terms with \(A{-}B\) mixing, as the
projection operators force it to be diagonal in the sublattice space. We assume that the radius of the flat band is much smaller than the maximum phonon momentum, so that the phonon cutoff does not need to be enforced in the momentum sum.

With a contact interaction, the only differences between the Hartree
and Fock terms come from the sign change (from the fermionic loop in
the Hartree term) and from the summation over spins. The total Coulomb self-energy is
\begin{equation}
\begin{split}
\Sigma_{\sigma,\rho}^c = - u \sum_{\vec p, m} G_{\bar{\sigma},\rho}(\vec p, i\omega_m).
\end{split}
\end{equation}
Note that the self-energy for up spin is determined from the propagator for the down spin and vice versa.

For electron-phonon interaction we only include the Fock term. The Hartree term vanishes because there is no \(p{=}0\) -phonon mediating the Hartree interaction.
\begin{equation}
\Sigma_{\sigma,\rho}^{\rm ph}(i\omega_n) = -\sum_{\vec p, m} \lambda_{nm} G_{\sigma,\rho}(\vec p, i\omega_m).
\end{equation}
This is written without particle-hole (Nambu) basis and must be
extended to include superconductivity. For now, we consider the normal
state to find what would be the stable phase if
superconductivity would not be present.

The total self-energy for spin \(\sigma\) and surface \(\rho\) is
\begin{equation}
\Sigma_{\sigma,s}(i\omega_n) = - \sum_{\vec p, m} \left[ \lambda_{nm} G_{\sigma,s}(\vec p, i\omega_m) + u G_{\bar\sigma,s}(\vec p, i\omega_m) \right].\label{eq:selfenergy1}
\end{equation}
Below we use this to study the possibility of a pseudospin analogue of the magnetic state. 

\section{Pseudospin magnetism}

Let us now consider the self-energy
\begin{equation}
\Sigma_{\rm ps}(i\omega_n) = -i\Sigma^\omega_n \mathds{1} + h^{\rm ps}_n \rho_3,\label{eq:se}
\end{equation}
where \(h^{\rm ps}\) is an analogous order parameter to ferromagnetic ordering, but with spin replaced by a pseudospin \(\rho\) (sublattice index). The propagator is 
\begin{equation}
G^{-1}(\vec p, i\omega_n) = i\tilde\omega_n\mathds{1} - \varepsilon_p \rho_1 - h^{\rm ps}_n \rho_3,
\end{equation}
where \(\tilde\omega_n = \omega_n + \Sigma^\omega_n = Z\omega_n\) is the renormalized frequency.

Instead of Eq.~\eqref{eq:selfenergy1} with a complicated matrix
structure, it is more useful to consider the components of the
self-energy that are symmetric and antisymmetric in spin and {sublattice indices \(\sigma\)} and \(\rho\). We get the self-consistency equations
\begin{align}
-i\Sigma^\omega_n &= -\sum_{\vec p, m} \lambda_{nm} \frac 1 4 \sum_{\sigma,\rho} G_{\bar\sigma,\rho}(\vec p, i\omega_m) + \Delta\mu,\\
h^{\rm ps}_n &= \sum_{\vec p, m} \left[ -u - \lambda_{nm}\right] \frac 1 4 \sum_{\sigma,\rho} \rho G_{\bar\sigma,\rho}(\vec p, i\omega_m),
\end{align}
where we isolate a correction to the chemical potential as
\(\Delta\mu\). We assume a fixed particle number, and therefore this
correction is counteracted by a shift in the chemical potential to the
opposite direction and as a result, it vanishes.

In the normal state, we can write \(G\) as a \(4\times4\) matrix. Its inverse is
\begin{equation}
\mathbf{G}^{-1}(\vec p, i\omega_n) = \mqty( 
i\tilde\omega_n - h^{\rm ps}_n & -\varepsilon_p & & \\
-\varepsilon_p & i\tilde\omega_n + h^{\rm ps}_n & & \\
& & i\tilde\omega_n - h^{\rm ps}_n & -\varepsilon_p \\
& & -\varepsilon_p & i\tilde\omega_n + h^{\rm ps}_n
),
\end{equation}
where the basis is chosen as \(\Psi^\dag_p = (c^\dag_{A\uparrow p},c^\dag_{B\uparrow p},c^\dag_{A\downarrow p},c^\dag_{A \downarrow p} )\). The matrix can be inverted in \(2\times2\) blocks labeled with spin:
\begin{equation}
\mathbf{G}_\sigma(\vec p, i\omega_n) = \frac{1}{\Omega_\sigma(\vec p, i\omega_n)}\mqty(
i\tilde\omega_n + h^{\rm ps}_n & \varepsilon_p \\
\varepsilon_p & i\tilde\omega_n - h^{\rm ps}_n
),
\end{equation}
where 
\begin{equation}
\Omega_\sigma(\vec p, i\omega_n) = (i\tilde\omega_n + h^{\rm ps}_n)(i\tilde\omega_n - h^{\rm ps}_n) - \varepsilon_p^2
= -\left[ \tilde\omega_n^2 + \varepsilon_p^2 + (h^{\rm ps}_n)^2 \right].
\end{equation}

The self-consistency equations for pseudo-spin magnetism are
\begin{align}
Z_n &= 1 + \sum_{\vec p, m} \lambda_{nm} \frac{\omega_m}{\omega_n}
\frac{Z_m}
{\tilde\omega_m^2 + \varepsilon_p^2 + (h^{\rm ps}_m)^2} \\
h^{\rm ps}_n &= \sum_{\vec p, m} \left( -u - \lambda_{nm} \right) \frac{h^{\rm ps}_m}{\tilde\omega_m^2 + \varepsilon_p^2 + (h^{\rm ps}_m)^2}.\label{eq:hps}.
\end{align}
We see from these equations that a pure pseudo-spin
magnetism can be ruled out, as the interactions do not support it:
both interaction terms in Eq.~\eqref{eq:hps} are negative; compare with Eqs.~(\ref{eq:scafm},\ref{sc:phi2}). We
would need to have a repulsive electron-phonon interaction or an
attractive Coulomb interaction in order to obtain a nonzero
solution. The unequal form of the interactions in spin-magnetism
versus pseudospin-magnetism originates from the lack of an exchange
term in the electron-phonon interaction.

\section{Antiferromagnetism}

For antiferromagnetism (AFM), the self-energy has the form
\begin{equation}
\Sigma_{\rm AFM}(i\omega_n) = -i\Sigma^\omega_n \mathds{1} + h_n \tau_3 \sigma_3, \rho_3,\label{eq:afmse2}
\end{equation}
where \(h_n\) is the frequency-dependent exchange field. As for the pseudospin magnetism above, the self-consistency equations for AFM are
\begin{align}
Z_n &= 1 + \sum_{\vec p, m} \lambda_{nm} \frac{\omega_m}{\omega_n}
\frac{Z_m}
{\tilde\omega_m^2 + \varepsilon_p^2 +h_m^2}.\label{eq:scafmZ},\\
h_n &= \sum_{\vec p, m} ( u {-} \lambda_{nm} ) \frac{h_m}{\tilde\omega_m^2 + \varepsilon_p^2 + h_m^2},\label{eq:scafm}.
\end{align}
In contrast to Eq.~\eqref{eq:hps}, the sign of \(u\) is now positive, and this makes the normal state unstable to the AFM phase. {Here we take the momentum cutoff to infinity. For large \(N\) the results do not depend on the cutoff provided it is larger than \(p_{\rm FB}\). For \(N\leq2\) the momentum integration diverges without a cutoff, so the results below do not apply for those cases.}

To numerically solve the above equations, we need to impose a cutoff \(\omega_{\rm max}\) in the Matsubara summation. To do that, we need to replace the Coulomb interaction with a
modified term using the pseudopotential trick
\cite{morel1962calculation}. The form of the pseudopotential depends on the details
of the self-consistency equation, so it has to be formulated
separately for different equations. The differences are in the
details, and the basic idea stays the same: in the Coulomb part of the
self-energy we divide the Matsubara summation to low-energy and
high-energy parts. Then we solve for the self-energy term and define a
new interaction constant, which takes into account the high-energy
contribution \cite{morel1962calculation}. The exact  form of the contribution from the high energy
sum is the part which differs between different equations. We
formulate equations in a form which is easy to solve
numerically.

For AFM equations, we can calculate the high energy part to be 
\begin{equation}
\begin{split}
\alpha
&= 2 T \sum_{\mathclap{\abs{\omega_n}>\omega_{\rm max}}} \int_0^\infty \frac{\dd{p} p}{p_{\rm FB}^2} \frac{1}{\omega_n^2 + \varepsilon_p^2} 
 = 2 T \left( \sum_{\omega_n} \mkern10mu-\mkern-10mu \sum_{\abs{\omega_n}<\omega_{\rm max}}\right) \int \frac{\dd{p} p}{p_{\rm FB}^2} \frac{1}{\omega_n^2 + \varepsilon_p^2} \\
%&= \frac{1}{2} \int_0^\infty \frac{\dd{x} x}{\varepsilon_x} \tanh(\frac{\varepsilon_x}{2T})
% - \frac{\pi}{N \sin(\pi/N)} T \sum_{\abs{\omega_n}<\omega_c} \frac{\left(\omega_n/\varepsilon_0\right)^{2/N}}{2\omega_n^2}\\
&= \frac{x_{\rm max}^{2-N}}{\varepsilon_0 (N-2)}
 + \frac{1}{\varepsilon_0} \int_0^{x_{\rm max}} \mkern-10mu\dd{x} x^{1-N} \tanh(\frac{\varepsilon_0 x^N}{2T})
 - \frac{\pi}{N \sin(\pi/N)} 2 T \sum_{\mathclap{\abs{\omega_n}<\omega_{\rm max}}} \frac{\left(\omega_n/\varepsilon_0\right)^{2/N}}{2\omega_n^2},
\end{split}\label{eq:pseudoalpha}
\end{equation}
where the cutoff \(x_{\rm max}\) is chosen so that \(\varepsilon(p_{\rm FB} x_{\rm max}) \gg
2T\). An argument larger than 4 is already large enough in order to
approximate the hyperbolic tangent by unity, so we can choose \(x_c =
(8T/\varepsilon_0)^{1/N}\). In total, the pseudopotential is 
\begin{equation}
u^{-} = \frac{u}{1 - u\alpha}.
\end{equation}
For a better accuracy at the strong coupling and low-temperature
regime, we also include the Coulomb part of the exchange field \(h_c\)  in
Eq. \eqref{eq:pseudoalpha}, so that \(\alpha\) depends
self-consistently on \(h_c\). The non-self-consistent \(\alpha\) given
above is sufficient for the calculation of \(T_C\).

With cutoff and a pseudopotential, Eqs.~(\ref{eq:scafmZ}-\ref{eq:scafm}) become
\begin{align}
h_n &= 2 T \sum_{\mathclap{|\omega_m|<\omega_{\rm max}}} ( u^- {-} \lambda_{nm} ) \int_0^\infty \frac{dp\,p}{p_{FB}^2} \frac{h_m}{\tilde\omega_m^2 + \varepsilon_p^2 + h_m^2},\label{eq:scafmZ2}\\
Z_n &= 1 + 2 T \sum_{\mathclap{|\omega_m|<\omega_{\rm max}}} \lambda_{nm} \frac{\omega_m}{\omega_n}
\int_0^\infty \frac{dp\,p}{p_{FB}^2} \frac{Z_m}
{\tilde\omega_m^2 + \varepsilon_p^2 +h_m^2},\label{eq:scafm2}
\end{align}
which can be solved numerically.

\subsection{Linearized equations for solving the critical temperature}

To determine the critical temperature, the self-consistency equations can be linearized with respect to \(h\). In this case the momentum integrals can also be done analytically. We obtain
\begin{align}
Z_n %&= 1 + T \sum_{\omega_m} \lambda_{nm} \frac{\omega_m}{\omega_n} \int_0^\infty \dd{x} 2 x \frac{Z_m}{\tilde\omega_m^2 + \varepsilon_x^2},\\
&= 1 + \alpha_N T \sum_{\mathclap{|\omega_m|<\omega_{\rm max}}} \frac{\lambda_{nm}}{\omega_n \tilde\omega_m} \left( \frac{\tilde\omega_m}{\varepsilon_0} \right)^{2/N},\nonumber\\
h_n %&= T \sum_{\omega_m} \left[ u^{-} - \lambda_{nm} \right] 
%\int_0^\infty \dd{x} 2x \frac{h}
%{\tilde\omega_n^2 + \varepsilon_x^2}\\
&= \alpha_N T \sum_{\mathclap{|\omega_m|<\omega_{\rm max}}} \left[ u^{-} - \lambda_{nm} \right]  \frac{h_m}{\tilde\omega_m^2}  \left( \frac{\tilde\omega_m}{\varepsilon_0} \right)^{2/N},\label{eq:afmsc}
\end{align}
where \(\alpha_N = N \sin(\pi/N)/\pi\).

If \(\lambda=0\), then \(Z=1\) and also the Matsubara summation can be
done analytically. We find \(\TCm\) in terms of the Riemann \(\zeta\)-function,
\begin{equation}
\TCm = \frac{1}{2\pi}\left[\frac{u\;\zeta(2{-}\frac{2}{N})\left(2^{2{-}\frac{2}{N}}{-}1\right)}{\varepsilon_0^{2/N} N \sin(\frac{\pi}{N})} \right]^{\frac{N}{N-2}},
\end{equation}
which approaches the value \(\TCm = u/4\) when \(N\to\infty\).

\section{Superconductivity}

Extending the formalism to the particle-hole space to include superconductivity, we define a Nambu vector \(\Psi^\dag = (\psi_{A,\vec p, \uparrow}^\dag, \psi_{B,\vec p,\uparrow}^\dag, \psi_{A,\vec p,\downarrow}^\dag, \psi_{B,\vec p,\downarrow}^\dag,\psi_{A,-\vec p,\uparrow}, \psi_{B,-\vec p,\uparrow}, \psi_{A,-\vec p,\downarrow}, \psi_{B,-\vec p,\downarrow}) \). 
The Feynman rules are then changed so that the interaction vertex gets an additional Nambu structure; \(P_\rho\) is replaced by \(P_\rho \tau_3\). The Hartree-Fock self-energy terms are
\begin{align}
\check\Sigma_{{\rm H}}^{\rm c} &= - u P_\rho \tau_3 \Tr[ P_\rho \tau_3\check G(\vec p, i\omega_n) ]
%= - u \tau_3 \sum_{P,\sigma,\tau} \tau G_{\sigma\sigma,\tau\tau,\rho\rho}
\\
\check\Sigma_{\rm F}^{\rm c} &= u \sum_{\vec p, m,\rho} P_\rho \tau_3 \check G(\vec p, i\omega_m) \tau_3 P_\rho\\
\check\Sigma_{\rm H}^{\rm ph} &= 0\\[0.7em]
\check\Sigma_{\rm F}^{\rm ph}(i\omega_n) &= - \sum_{\vec p, m,\rho} \lambda_{nm} P_\rho \tau_3 \check G(\vec p, i\omega_m) \tau_3 P_\rho.
\end{align}
We note that the Hartree term only affects the normal-state self-energy components, and not the off-diagonal ones associated with superconductivity. Superconductivity is determined only from the Fock terms.

The order parameter is the same on both sublattices, \(\phi_A =
\phi_B\). In principle, the order parameter could also have a
different phase and magnitude on the two surfaces, but this choice is
the one with the lowest energy \cite{Kauppila16}. The self-consistency equations for the superconducting phase are
\begin{align}
Z_n = 1 + 2 T\sum_{\omega_m} \lambda_{nm} \frac{\omega_m}{\omega_n} \int_0^\infty \frac{\dd{p}  p}{p_{\rm FB}^2} \frac{Z_m}{\tilde\omega_m^2 + \varepsilon_p^2 + \phi_m^2},\label{sc:phi1}\\
\phi_n = 2 T\sum_{\omega_m} ( \lambda_{nm} - u ) \int_0^\infty \frac{\dd{p} p}{p_{\rm FB}^2} \frac{\phi_m}{\tilde\omega_m^2 + \varepsilon_p^2 + \phi_m^2}.\label{sc:phi2}
\end{align}

In Eq.~\eqref{sc:phi2} we can impose a Matsubara cutoff \(\omega_{\rm max}\) if we simultaneously replace \(u\) with a pseudopotential \(u^{+}\), like in
Eqs.~(\ref{eq:pseudoalpha}, \ref{eq:scafm2}). Compared to the AFM case, there is a sign change in the pseudopotential,
\begin{equation}
u^{+} = \frac{u}{1+u\alpha}.
\end{equation}
The interpretation of the pseudopotential is that for superconductivity, the scattering at high energies reduces the effect of interaction at lower energies, whereas for magnetism, the effect is reversed. If we remove the high-energy scattering from the theory, the interaction has to be replaced by an effective pseudopotential to account for their effect.

%To solve for the critical temperature, Eqs.~\eqref{sc:phi1} and \eqref{sc:phi2} can be linearized with respect to \(\phi\). The resulting equations have the same form as in the AFM case.

\subsection{Correction to the critical temperature due to retardation}

To obtain an approximation for the effect of the electron-phonon renormalization on superconductivity, we first solve the renormalization function \(Z\) at zero temperature by approximating the self-consistency equation as
\begin{equation}
Z(i\omega) = 1 + \lim_{\epsilon\to 0} \frac{\lambda}{\omega} \int \frac{\dd{\omega'}}{2\pi} \frac{\omega_E^3}{\omega_E^2 + (\omega-\omega')^2 } \frac{Z_0 \omega' }{(Z_0 \omega')^2 + \epsilon^2},
\end{equation}
where \(\epsilon=0^+\) is used to regularize the integral. Inside the
integral we approximate \(Z\) by its peak value \(Z_0\equiv
Z(i\omega=0)\). The integral yields
\begin{equation}
Z(i\omega) = 1 + \frac{\lambda\omega_E^2}{Z_0 (\omega_E^2 + \omega^2)}.
\end{equation}
For \(\omega=0\), we have \(Z_0 = 1 + \lambda/Z_0\), whose solution in the first order in \(\lambda\) is \(Z_0 = 1 + \lambda\).

Now the approximate self-consistency equation for \(\phi\) is
\begin{equation}
\phi_0 = \lambda \int \frac{\dd{\omega}}{2\pi} \frac{\omega_E^3}{\omega_E^2 + \omega^2} \frac{\phi_0}{(Z\omega)^2 + \phi_0^2} \xrightarrow[\text{}\phi_0 \to 0]{} \frac{\lambda\omega_E}{2 Z_0}.
\end{equation}

At zero temperature, for \(\Delta=\phi/Z\), we have
\begin{equation}
\Delta_0 = \frac{\phi_0}{Z_0} = \frac{\lambda \omega_E}{2 Z_0^2} =  \frac{\lambda \omega_E}{2(1+2\lambda)}.\label{eq:DeltaMcMillan}
\end{equation}
At weak coupling, the critical temperature is half the value of \(\Delta_0\),
\begin{equation}
\TCsc \approx \frac{\Delta_0}{2} =  \frac{\lambda \omega_E}{4(1+2\lambda)}.\label{eq:TCMcMillan}
\end{equation}
The above calculation accounts for the electron-phonon renormalization
in a very crude manner. A better agreement with the numerical results
is obtained by a direct fit to the numerical results. This way we obtain {Eq.~(15)} in the main text.

\subsection{Linear dispersion}

An alternative to the electronic Hamiltonian with the \(p^N\)-dispersion used in the main paper is a Dirac Hamiltonian
\begin{equation}
H_{{\rm el}, p} = \vF
\begin{pmatrix}
0 & p_x - i p_y \\ p_x + i p_y &0
\end{pmatrix},
\label{eq:linear2}
\end{equation}
where \(\vF\) is the Fermi velocity. The pseudospin-structure of the Hamiltonian is unchanged from Eq.~(2) in the main text and it can also be considered as the \(N=1\) case of the \(p^N\)-dispersion. Unlike for the \(p^N\)-dispersion with $N>2$ for which we can approximate the cutoff as infinite, the results for the linear dispersion are highly dependent on the cutoff. We assume a momentum cutoff \(p_c\) and approximate the Brillouin zone as being circular. The momentum cutoff corresponds to energy cutoff \(\varepsilon_c = v_F p_c\).

This Hamiltonian is realized approximately in twisted bilayer graphene (TBG) when the twist angle between the two graphene layers is larger than the magic twist angle of \(\theta_{\rm magic}\approx1.1\degree\). TBG is only periodic in a large scale moir\'e superlattice, corresponding to a superlattice Brillouin zone which is small compared to graphene Brillouin zone. The lowest energy bands are described by the Hamiltonian \eqref{eq:linear2} with \(v_F\) becoming smaller and smaller as the twist angle approaches the magic angle. The rest of the Brillouin zone of the original graphene is folded into higher energy bands which we neglect. The low-energy Hamiltonian for TBG can be derived with perturbation theory when the twist angle \(\theta > 1.8\degree\) [\onlinecite{dos2012continuum}]. In the perturbative model, the layers are decoupled from each other. The simple perturbation expansion fails near the magic angle, but the layers remain decoupled up to the magic angle \cite{morell2010flat}. 

We describe TBG in terms of bare graphene, so in applying Eq. (9) of the main paper we take $\Omega_{\rm FB}$ as the size of the superlattice Brillouin zone and $\Omega_{\rm BZ}$ as the original Brillouin zone of graphene. In the momentum integrals the cutoff \(p_c\) is the radius of the superlattice Brillouin zone $p_{FB}$. The interaction constants become \(\lambda = g^2/n\omega_E^2\) and \(u = U/n\), where \(n\) is the ratio between the areas of the original graphene and the superlattice Brillouin zones, or equivalently the ratio between the number of lattice sites in the superlattice and the original graphene unit cells.

After doing the momentum integrals, the self-consistency equations for superconductivity become
\begin{align}
\phi_n &= T \sum_{\mathclap{m=-\infty}}^\infty ( \lambda_{nm} {-} u ) \frac{\phi_m}{\varepsilon_c^2} \log\Big[1+\frac{\varepsilon_c^2}{\tilde\omega_m^2 + \phi_m^2}\Big],\label{eq:eliashberg_sc_phi_lineardisp}\\
Z_n &= 1 + T \sum_{\mathclap{m=-\infty}}^\infty \lambda_{nm} \frac{\omega_m}{\omega_n}
\frac{Z_m}{\varepsilon_c^2} \log\Big[1+\frac{\varepsilon_c^2}{\tilde\omega_m^2 + \phi_m^2}\Big],\label{eq:eliashberg_sc_Z_lineardisp}
\end{align}
Again, antiferromagnetism has a similar set of equations. If we assume that \(\varepsilon_c \ll \phi_0\) and \(\varepsilon_c \ll \omega_E\) so that the argument of the logarithm is small for both small and large \(\omega_m/\omega_E\), the logarithm can be approximated with the first order term,
\begin{align}
\frac{1}{\varepsilon_c^ 2}\log\Big[1+\frac{\varepsilon_c^2}{\tilde\omega_m^2+\phi_m^2}\Big] \approx \frac{1}{\tilde\omega_m^2+\phi_m^2}.
\end{align}
In this limit, the dependence on the cutoff energy vanishes and we recover the self-consistency equations for a completely flat band.

Unlike for the \(p^N\) dispersion, superconductivity only appears for the linear dispersion if \(\lambda\) is stronger than some critical value \(\lambda_{C,0}\) even if \(u=0\) [\onlinecite{kopnin2008bcs}]. This is due to vanishing density of states near zero energy. This value can be found by linearizing the self-consistency equation \eqref{eq:eliashberg_sc_phi_lineardisp} both in \(\phi\) and \(\TCsc\). For an interaction without retardation, we find that \(\lambda_{C,0}=\varepsilon_c/\omega_E\). With \(\varepsilon_c \ll \omega_E\), the phase diagram of Fig. 5 in the main paper for the linear case is the same as for \(N=\infty\), except near the phase boundary, where the normal state is the ground state.

\subsection{Effect of Debye dispersion}
In a real material, the phonon dispersion is obviously not described
by the Einstein model adopted in the text. To get an idea how much the
exact phonon dispersion affects the results, we consider also the
Debye model.

For the Einstein model phonons there is no momentum dependence in the interactions, and for this reason the self-energy is also independent of momentum. For Debye phonons the interaction obtains a momentum dependence through the phonon dispersion. Without calculating the full momentum dependent theory, we can estimate the effect of the dispersion by estimating the typical energy of the exchanged phonon and the average interaction constant.

The maximum phonon energy exchanged within the flat band is limited by
the flat band diameter \(2p_{\rm FB}\) to be \(\omega_0 = {p_{\rm
    FB}\omega_D}/{q_M}\), where \(\omega_D\) is the Debye energy and
\(q_M\) is the maximum phonon momentum. As typically \(q_M\) is of the
order of the size of the Brillouin zone, \(\omega_0 \ll \omega_D\) and the energy scale is reduced. On the other hand, we find that the dimensionless interaction constant is enhanced: \(\lambda \,{\propto}\,1/\omega_0^2\). However, because \(\Delta_0 \propto\lambda^{0.2} \omega_0\), the magnitude of \(\Delta_0\) is restricted by \(\omega_0\) and the total effect is a smaller critical temperature than with the Einstein phonons with energy \(\omega_E\) equal to the Debye energy.

The above concerns interactions within one flat band. We can also consider Debye phonons in the context of two flat bands separated by the distance \(p_d\gg p_{FB}\) in momentum space. The momentum range of Debye phonons connecting the two parts of the Brillouin zone is limited to \(p_d\), and they can be treated as if they had a constant energy \(\omega_0 = \omega_D p_d/q_M\). In this case they act essentially as Einstein phonons with an effective energy scale \(\omega_0\).

% The states at the edge of the flat band can exchange higher energy phonons than the states at the middle of the flat band. This makes the order parameter inhomogeneous within the flat band. Assuming that the states outside the flat band are not within the phonon energy range we can estimate the interaction by averaging it over the flat band.  The average interaction strength \(\lambda\) on the  other hand, is increased, as it can be shown to scale inversely with \(\omega_0\).

\section{Equations for competing magnetic and superconducting phases}

We now concentrate on the possible coexistence of antiferromagnetism
and superconductivity for $N\rightarrow \infty$. The presence of AFM (\(\rho_3\sigma_3\tau_3\)) and the singlet superconducting order parameter (\(\sigma_2\tau_2\)) induces a triplet component proportional to
\begin{equation}
\rho_3 \sigma_3 \tau_3 \times \sigma_2\tau_2 \propto \rho_3 \sigma_1\tau_1.
\end{equation}
The triplet is induced into the propagator, but it also enters in the
self-energy if the interactions support it. This requires that the
interactions have an odd-frequency part, which is true with the
retarded interaction, but not present with the instantaneous interaction.

The inverse propagator is
\begin{equation}
\check G^{-1}(i\omega_n) = i\tilde\omega_n \mathds{1} - h  \rho_3 \sigma_3 \tau_3 - \phi \sigma_2 \tau_2 - i d \rho_3 \sigma_1 \tau_1.
\end{equation}
To invert this, we notice that the matrix separates into four \(2\times2\) blocks. Labeling these blocks by the spin and pseudo-spin of their particle part (the hole part is associated with the opposite spin and same pseudo-spin), we can write them as
\begin{equation}
G^{-1}_{\rho\sigma}(i\omega_n) = 
\mqty(
	i\tilde\omega_n - \rho \sigma h_n & -\sigma \phi_n + i \rho d_n \\
	-\sigma \phi_n + i \rho d_n & i\tilde\omega_n - \rho \sigma h_n
).
\end{equation}
The inverse of this is
\begin{equation}
\begin{split}
G_{\rho\sigma}(i\omega_n) &=
-\frac
{(\tilde\omega_n^2-h_n^2+\phi_n^2-d_n^2)-2i \rho \sigma (h_n\tilde\omega_n+\phi_n d_n)}
{(\tilde\omega_n^2-h_n^2+\phi_n^2-d_n^2)^2 + 4(h_n\tilde\omega_n + \phi_n d_n)^2}
\mqty(
	i\tilde\omega_n - \rho \sigma h_n & \sigma \phi_n - i \rho d_n \\
	\sigma \phi_n - i \rho d_n & i\tilde\omega_n - \rho\sigma h_n
)\\
&\equiv
-\frac
{\gamma_n-i \rho \sigma \delta_n}
{\zeta_n}
\mqty(
	i\tilde\omega_n - \rho \sigma h_n & \sigma \phi_n - i \rho d_n \\
	\sigma \phi_n - i \rho d_n & i\tilde\omega_n -  \rho\sigma h_n
)\\
&=
-\frac{i}{\zeta_n}\mqty( \gamma_n\tilde\omega_n + \delta_n h_n & -\rho_n( \gamma_n d_n + \delta_n \phi_n ) \\ -\rho_n( \gamma_n d_n + \delta_n \phi_n ) & \gamma_n\omega_n + \delta_n h_n )
+ \frac{1}{\zeta_n}\mqty(\rho\sigma(\gamma_n h_n - \delta_n \tilde\omega_n) & -\sigma( \gamma_n\phi_n - \delta_n d_n )
\\ -\sigma( \gamma_n\phi_n - \delta_n d_n ) & \rho \sigma(\gamma_n h_n - \delta_n \tilde\omega_n)),
\end{split}
\end{equation}
where, in the last line, we have separated the odd and even frequency
parts. Above, we define
\begin{align}
\gamma_n &= \tilde\omega_n^2-h^2_n+\phi^2_n-d^2_n,\\
\delta_n &= 2(h_n\tilde\omega_n+\phi_n d_n),\\
\zeta_n &= \gamma^2_n + \delta^2_n.
\end{align}
From these, \(\tilde\omega\), \(d\) and \(\delta\) are odd in \(\omega_n\), and the other terms are even in \(\omega_n\). The diagonal self-energies depend on the propagator for the inverted spin, and for them we add a sign change to \(h\) and \(\phi\).

The self-consistency equations are
\begin{align}
d_n &= T \sum_{\mathclap{|\omega_m|<\omega_{\rm max}}} \lambda_{nm}^- \frac{\gamma_m d_m + \delta_m \phi_m}{\zeta_m},\label{eq:ce_sc_d}\\
\Sigma^\omega_n &= -T\sum_{\mathclap{|\omega_m|<\omega_{\rm max}}} \lambda^-_{nm} \frac{\gamma_m \tilde\omega_m + \delta_m h_m}{\zeta_m},\label{eq:ce_sc_sigma}\\
\phi_n &= T \sum_{\mathclap{|\omega_m|<\omega_{\rm max}}} \left[ \lambda^+_{nm} - u_c^{+} \right]  \frac{\gamma_m \phi_m - \delta_m d_m}{\zeta_m},\label{eq:ce_sc_phi}\\
h_n &= T\sum_{\mathclap{|\omega_m|<\omega_{\rm max}}} \left[ \lambda^+_{nm} - u_c^{-}\right] \frac{\gamma_m h_m - \delta_m \tilde\omega_m}{\zeta_m}\label{eq:ce_sc_h},
\end{align}
where we assume the Matsubara sum to have a cutoff which determines the values of the pseudopotential terms. The equation for \(\Sigma^\omega\) can also be expressed in terms of \(Z\) as
\begin{equation}
Z_n = 1 + T\sum_{\mathclap{|\omega_m|<\omega_{\rm max}}} \lambda^-_{nm} \frac{\gamma_m \tilde\omega_m + \delta_m h_m}{\zeta_m}.
\end{equation}

The odd and even electron-phonon interaction kernels are
\begin{equation}
\lambda^\pm_{nm} = \frac 1 2 \left[\lambda(\omega_n-\omega_m) \pm \lambda(\omega_n+\omega_m) \right] \omega_E.
\end{equation}
The Coulomb interaction only has an even part, so it only affects the even-frequency self-energy terms, namely \(\phi\) and \(h\). The coexistence modifies the pseudopotentials \(u^{\pm}\) slightly as the high-frequency part of both order parameters has to be taken into account. It is still defined as \(u^\pm = u/(1\pm u_c \alpha_\pm)\), but now with
\begin{equation}
\alpha_\pm = \frac{\sinh(\frac{\phi_c+h_c \pm (\phi_c - h_c)}{2T})}{(\phi_c+h_c \pm (\phi_c - h_c))\left[\cosh(\frac{\phi_c}{T})+\cosh(\frac{h_c}{T})\right]}
- T\sum_{\mathclap{|\omega_m|<\omega_{\rm max}}} \frac{\omega_m^2 \mp h_c^2 \pm \phi_c^2}{(\omega_n^2 - h_c^2 + \phi_c^2)^2 + 4 h_c^2 \omega_n^2}.
\end{equation}
We solve these equations numerically to study the competition
  of the two phases.

\section{A toy model for the competition between the phases}

\begin{figure*}
	\subfloat[\label{fig:free}]{% 
    	\includegraphics[width=0.4\textwidth]{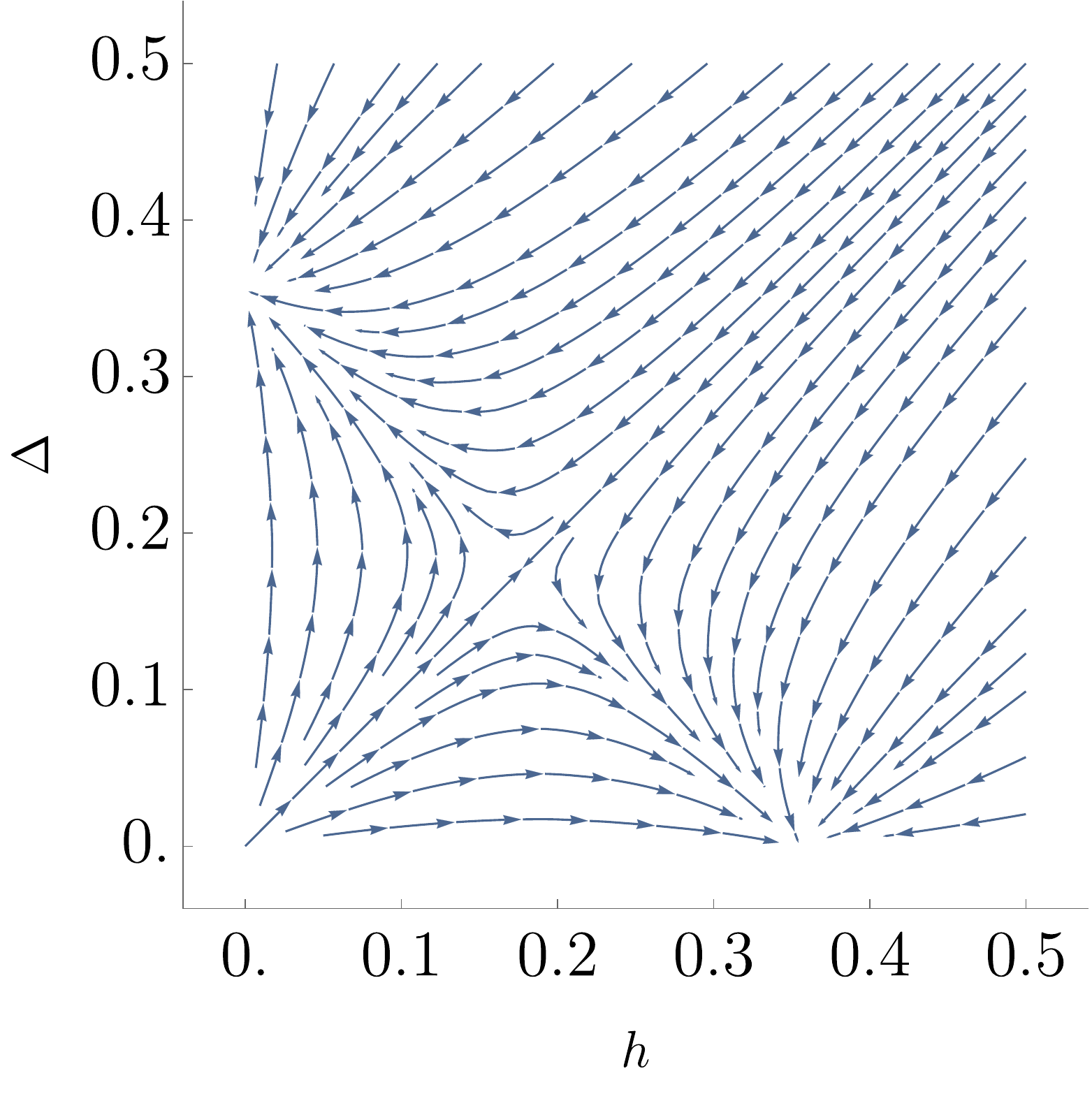}   
  } 
  \hfill
  	\subfloat[\label{fig:phasediagram2}]{% 
    	\includegraphics[width=0.4\textwidth]{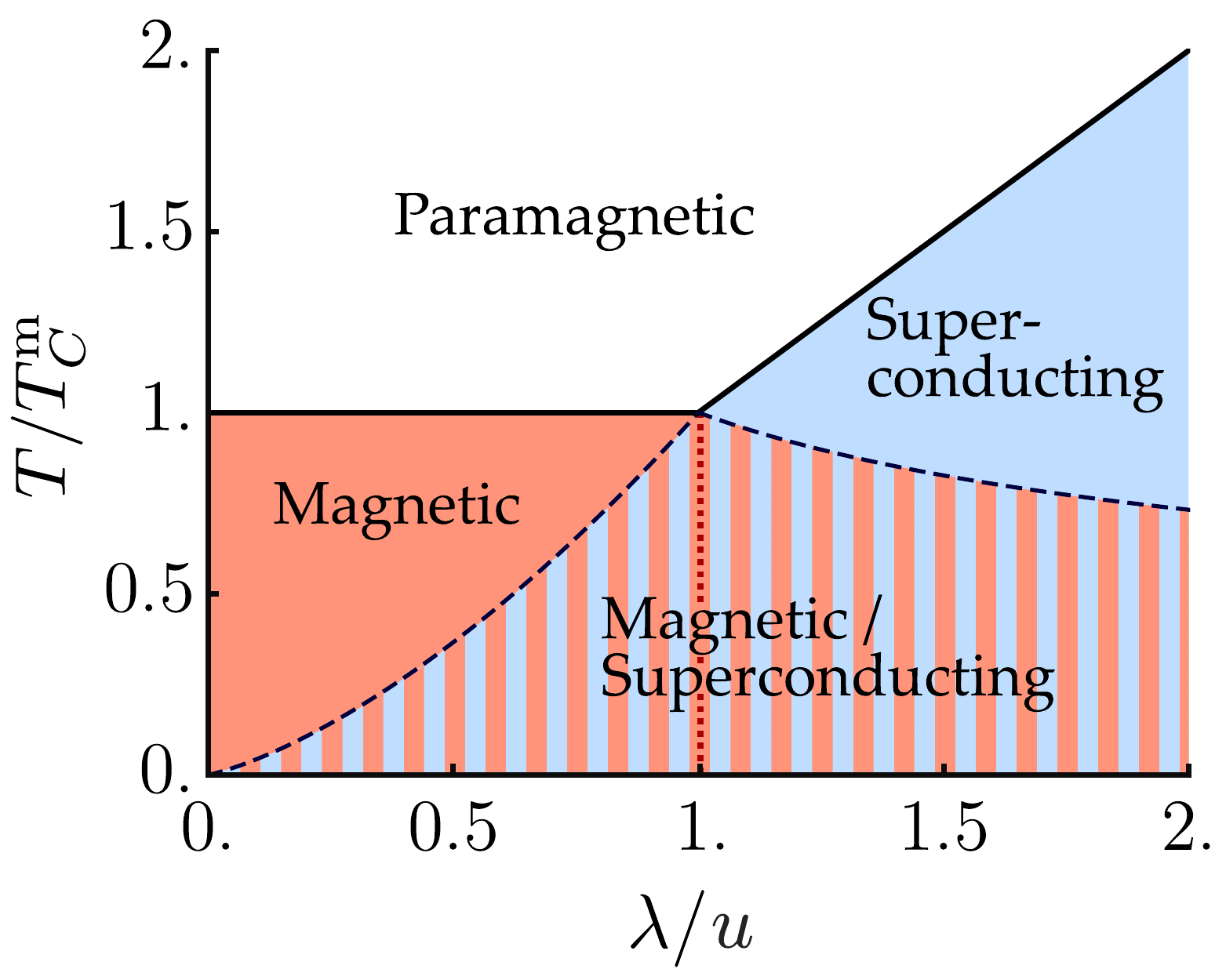}
  } 
  \caption{(a) Stream plot of the gradient of the free energy
    \eqref{eq:free} for \(\lambda=u\) at temperature \(T<\TCsc,
    \TCm\). There are two local minima, which correspond to the
    magnetic and superconducting phases, and a saddle point which is
    the unstable coexistence solution. (b) Phase diagram with fixed \(u\). When \(\lambda<u\) and \(T<\TCm\) , the thermodynamically stable state is
    the magnetic state and when \(\lambda>u\) and and \(T<\TCsc\) the stable state is the superconducting state. The dividing line \(\lambda=u\) is marked with the red dotted line. On the striped region, both phases are possible as metastable states.}
\end{figure*}

To understand the phenomenology of the two co-existing order
parameters, let us consider a toy model where there are two order
parameters \(\Delta\) and \(h\), and two separate interaction
channels, \(\lambda\) and \(u\). We assume that the interactions are
separate in the sense that \(\lambda\) only mediates superconductivity
and \(u\) mediates the magnetization. For simplicity, we assume that the interactions are frequency independent so there is no triplet component and no electron-phonon renormalization term.

This model is not realized in the weak coupling as in that case the interaction channels are not separate, but have always opposing signs. The superconducting channel is mediated by the interaction with strength \(\lambda\omega_E-u\) and the magnetic channel has the strength \(u-\lambda\omega_E\). The channels are separate only in strong coupling, as the low interaction frequencies are attractive to superconductivity and the high frequencies for magnetism.

In the the full coexistence model the dispersions for two non-equivalent bands are \(E_{p} = \Delta \pm \sqrt{\varepsilon_p^2 + h^2}\). For \(N\to\infty\) this reduces to \(E_{p\sigma} = \Delta \pm h\). We take this dispersion as the starting point. For notational simplicity, we assume that \(\Delta>0\) and \(h>0\). The fields \(\Delta\) and \(h\) are also included in the partition function,
\begin{equation}
\begin{split}
\frac{Z}{Z_0} = Z_0^{-1}&\prod_{\mathclap{p\in \rm{FB}, \omega_n}} \left[\omega_n^2 + (\Delta + h)^2\right] \times \left[\omega_n^2 + (\Delta - h)^2\right] \times e^{-\beta \Delta^2/\lambda} \times e^{-\beta h^2/u}\\
= &\prod_{p\in \rm{FB}} \cosh(\frac{\beta(\Delta + h)}{2}) \cosh(\frac{\beta(\Delta - h)}{2}) \times e^{-\beta \Delta^2/\lambda} \times e^{-\beta h^2/u}.
\end{split}
\end{equation}
Above, the product over the Matsubara frequencies is evaluated using a standard Matsubara trick. The overall constant cancels against the normal state partition function \(Z_0\). The free energy relative to the normal state is 
\begin{equation}\label{eq:free}
\begin{split}
F(\Delta, h) = -T \log(\frac{Z}{Z_0}) &= C\left( \frac{\Delta^2}{\lambda} + \frac{h^2}{u} - T \log\left[\cosh(\frac{\Delta+h}{2T}) \cosh(\frac{\Delta-h}{2T}) \right] \right) \\
&= C\left( \frac{\Delta^2}{\lambda} + \frac{h^2}{u} - T \log\left[ \frac 1 2\cosh(\frac{\Delta}{T}) + \frac 1 2 \cosh(\frac{h}{T}) \right] \right),
\end{split}
\end{equation}
where the momentum sum gives the multiplicative factor \(C>0\).

The self-consistency equations are given as derivatives of the free energy with respect to fields \(\Delta\) and \(h\):
\begin{align}
\Delta &= \frac{\lambda}{2} \frac{\sinh(\Delta/T)}{ \cosh(\Delta/T) + \cosh(h/T)},\label{eq:sc1}\\
h &= \frac{u}{2} \frac{\sinh(h/T)}{ \cosh(h/T) + \cosh(\Delta/T)}.\label{eq:sc2}
\end{align}

As a check, we see that if there was no frequency dependence in the interactions, Eqs.~\eqref{eq:ce_sc_phi} and \eqref{eq:ce_sc_h} would give similar equations after the Matsubara summation.
%\begin{align}
%T\sum_{\omega_m} \frac{2 (\omega_m^2 - h^2 + \Delta^2) \Delta}{(\omega_m^2 - h^2 + \Delta^2)^2 + 4h^2 \omega_m^2} = \frac{\sinh(\Delta/T)}{ \cosh(\Delta/T) + \cosh(h/T)},\\
%T\sum_{\omega_m} \frac{2(\omega_m^2 + h^2 - \Delta^2) h}{(\omega_m^2 - h^2 + \Delta^2)^2 + 4h^2 \omega_m^2} = \frac{\sinh(h/T)}{ \cosh(h/T) + \cosh(\Delta/T)}.
%\end{align}
%
The difference is that in the full model with instantaneous interactions, there is effectively only one interaction constant \(\lambda_{\rm eff} = \lambda\omega_E {-} u\).

The self-consistency equations \eqref{eq:sc1} and \eqref{eq:sc2} for \(\Delta\) and \(h\) should be solved simultaneously. At low temperatures, \(T\ll h,\Delta\), we can approximate the hyperbolic functions with exponentials, and obtain
\begin{equation}
\Delta(T=0) = \lim_{T\to 0} \frac{\lambda}{2[1+\exp(\frac{h-\Delta}{T})]} = 
\begin{cases}
\Delta_0 &\text{for } h < \Delta\\
\Delta_0/2 &\text{for } h = \Delta\\
0 &\text{for } h > \Delta,
\end{cases}\label{eq:zeroDelta}
\end{equation}
where \(\Delta_0 = \lambda/2\) is the zero-temperature order parameter for \(h=0\). For \(h\) in terms of \(\Delta\), an analogous expression can be found,
\begin{equation}
h(T=0) = \lim_{T\to 0}\frac{u}{2[1+\exp(\frac{\Delta-h}{T})]} = 
\begin{cases}
h_0& \text{for } \Delta < h\\
h_0/2& \text{for } \Delta = h\\
0& \text{for } \Delta > h,
\end{cases}\label{eq:zeroH}
\end{equation}
with \(h_0=u/2\). At zero temperature, coexistence is only possible if
\(u=\lambda\) and \(\Delta=h=\lambda/4\). This coexistence point is a
saddle point of free energy. This is illustrated in
Fig.~\ref{fig:free}. Similar kind of solutions are also found at a finite temperature.

Assuming a second order phase transition from the normal state to a superconducting state, we can linearize \eqref{eq:sc1} with \(h=0\) to find a critical temperature \(\TCsc = \lambda/4\). We will see below that when \(u>\lambda\) the phase transition is actually of the first order from a magnetic state to the superconducting state, but we still use the above definition for \(\TCsc\) to set a temperature scale. Similarly, assuming a second order phase transition from normal state to a magnetic state, we linearize \eqref{eq:sc2} to find a critical temperature \(\TCm=u/4\), which we take as the definition of \(\TCm\).

\subsection{Stability of the phase with lower \(T_C\)}

From the free energy we see that when \(\lambda < u\), the magnetic phase is stable and the superconducting phase is either metastable or unstable. When \(\lambda>u\), the roles are reversed. The solution is (meta)stable if it is a local minimum of the free energy. The criterion is
\begin{equation}
\frac{\partial^2 F }{\partial h^2}\frac{\partial^2 F }{\partial \Delta^2} - \left(\frac{\partial^2 F }{\partial \Delta\partial h}\right)^2 >0 \quad\text{and}\quad \frac{\partial^2 F }{\partial h^2}>0.
\end{equation}
When \(h=0\) or \(\Delta=0\) the cross derivatives vanish, and the stability condition becomes \({\partial^2 F }/{\partial h^2}>0\) and \({\partial^2 F }/{\partial \Delta^2}>0\).

Let us consider the (meta)stability of the superconducting phase when \(0<\lambda<u\) and \(T<\TCsc\), where \(\TCsc=\lambda/4\). The stability condition becomes
\begin{align}
0 < \left.\frac{\partial^2 F}{\partial h^2}\right|_{h=0} = C \left( \frac{2}{u} - \frac{1}{T+T\cosh(\frac{\Delta}{T})} \right).
\end{align}
At low temperatures, the first term in the parentheses dominates and the superconducting phase is metastable. Near \(\TCsc\) the second term dominates and the superconducting phase becomes unstable against a spontaneous magnetization. In this case, \(\Delta\approx0\), and 
\begin{equation}
\frac{2}{u} - \frac{1}{T+T\cosh(\frac{\Delta}{T})} \approx \frac{2}{u} - \frac{1}{2 \TCsc} = 2\left(\frac{1}{u}-\frac{1}{\lambda}\right) < 0.\label{eq:criterion}
\end{equation}
The transition temperature \(T^*\) at which the system becomes unstable can be determined from the condition \({\partial^2 F }/{\partial h^2}=0\), which is equivalent to solving the magnetic critical temperature by linearizing Eq. \eqref{eq:sc1} with \(\Delta\) as solved from Eq.~\eqref{eq:sc2} with \(h=0\).

For \(\lambda > u > 0\), the magnetic phase is the metastable one, and the equations apply after interchanging \(u\leftrightarrow\lambda\), \(h\leftrightarrow\Delta\) and \(\TCsc\leftrightarrow\TCm\). The numerical solution for the transition temperature of both phases is shown in Fig.~\ref{fig:phasediagram2} as the dashed line.

Similar stability analysis can be used with the full model, except in that case the frequency dependence of the self-energy functions complicates the situation. We can however simplify the problem by projecting the order parameters on the linearized solution.

\section{Competition in the full frequency-dependent model}

Now we return to the full frequency-dependent Eqs. \eqref{eq:ce_sc_d}--\eqref{eq:ce_sc_h}. We do for the full model the same kind of analysis as in the previous section for the toy model. As explained after \eqref{eq:criterion}, the stability is determined by solving the critical temperature of the phase with higher \(T_C\) in the presence of the other order parameter.

\subsection{Stability of the superconducting phase}

We now consider parameters \(\lambda, u\) to be such that \(0 <
\TCsc < \TCm\) and determine the temperature \(T^*\) above which
the superconducting phase is unstable against a magnetic instability.

To do this, we linearize Eqs.~\eqref{eq:ce_sc_d}--\eqref{eq:ce_sc_h}
with respect to \(h\) and \(d\). We assume the singlet order parameter \(\phi\) to be finite. The equations for \(\Sigma^\omega\) and \(\phi\) are
\begin{align}
\Sigma^\omega_n &= -T\sum_{\mathclap{|\omega_m|<\omega_{\rm max}}} \lambda^-_{nm} \frac{\tilde\omega_m}{\tilde\omega_m^2 + \phi_m^2},\\
\phi_n &= T\sum_{\mathclap{|\omega_m|<\omega_{\rm max}}} (\lambda^+_{nm} - u^+ )\frac{\phi_m}{\tilde\omega_m^2 + \phi_m^2}.
\end{align}

The equations for \(h\) and \(d\) are coupled to one \(2\times2\) matrix,
\begin{equation}
\mqty[d_n\\ h_n] = T\sum_{\mathclap{|\omega_m|<\omega_{\rm max}}} \frac{1}{(\tilde\omega_m^2 + \phi_m^2)^2} \mqty[\lambda_{nm}^-(\tilde\omega_m^2+3\phi_m^2) & 2\lambda_{nm}^-\phi_m\tilde\omega_m \\ 2(u^- - \lambda^+_{nm})\phi_m\tilde\omega_m & (u^- - \lambda^+_{nm})(\tilde\omega_m^2 - \phi^2_m)] \mqty[d_m\\ h_m].
\end{equation}
With the high-frequency cutoff, the equation can be written as a
\(2M\times2M\) matrix, where \(M\) is the number of Matsubara
frequencies below the cutoff. The critical temperature \(T^*\) of this
AFM/triplet phase is then determined numerically by finding the
temperature at which the largest eigenvalue of the matrix becomes
larger than unity \cite{matsumoto2012coexistence}.

\subsection{Stability of the magnetic phase}

Conversely, let us now consider parameters \(\lambda, u\) such that
\(0 < \TCm < \TCsc\) and determine the temperature \(T^*\) above
which the magnetic phase is unstable against a superconducting
instability. Now we linearize
Eqs.~\eqref{eq:ce_sc_d}--\eqref{eq:ce_sc_h} with respect to \(\phi\)
and \(d\). We assume the field \(h\) to be finite. The equations for \(\Sigma^\omega\) and \(h\) are
\begin{align}
\Sigma^\omega_n &= -T\sum_{\mathclap{|\omega_m|<\omega_{\rm max}}} \lambda^-_{nm} \frac{\tilde\omega_m}{\tilde\omega_m^2 + h^2_m},\\
h_n &= T\sum_{\mathclap{|\omega_m|<\omega_{\rm max}}} \left[ u^- - \lambda^+_{nm} \right] \frac{h_m}{\tilde\omega_m^2 + h^2_m}.
\end{align}

The equations for \(\phi\) and \(d\) are coupled, and can be written as a matrix equation
\begin{equation}
\mqty[d_n\\\phi_n] = T\sum_{\mathclap{|\omega_m|<\omega_{\rm max}}} \frac{1}{(\tilde\omega_m^2 + h^2_m)^2} \mqty[
 \lambda^-_{nm} (\tilde\omega_m^2-h^2_m) &  -2\lambda^-_{nm} h_m\tilde\omega_m \\
 2 ( \lambda^+_{nm} -u^+ ) h_m\tilde\omega_m &  (\lambda^+_{nm} -u^+) (\tilde\omega_m^2 - h^2_m)
] \mqty[ d_m\\ \phi_m]
\end{equation}
Again, with the Matsubara cutoff, this is a matrix equation and the critical temperature \(T^*\) can be solved by searching for the temperature at which the largest eigenvalue crosses 1. The solution curve for both cases, \(\TCm < \TCsc\) and \( \TCm > \TCsc\), is shown as a dashed boundary line in the phase diagram, Fig. 1 in the main text.

\end{document}